\documentclass{aa}

\usepackage{graphicx}
\usepackage{txfonts}
\usepackage{hyperref}
\usepackage{amsmath}

\newcommand{\idefix}{\textsc{Idefix}}

\begin{document}

   \title{A nonrelativistic radiative transfer module for \idefix}

   \author{Nicolas Scepi\inst{1} \and Geoffroy Lesur\inst{1}}
 \authorrunning{N. Scepi et al.}
   \institute{Univ. Grenoble Alpes, CNRS, IPAG, 38000 Grenoble, France\\
 \email{nicolas.scepi@univ-grenoble-alpes.fr}}

   \date{Received September 15, 1996; accepted March 16, 1997}

  \abstract 
   {Radiation magnetohydrodynamic (RMHD) simulations are essential for comparisons with observations, particularly in the regime where fluids and radiation are dynamically coupled. Although computationally expensive, RMHD is becoming increasingly accessible with the advent of exascale computing. However, only a few public RMHD codes are currently able to fully exploit the diversity of modern accelerated architectures.}
   {We present a nonrelativistic radiative transfer module for the public magnetohydrodynamic code \idefix; it is built on the Kokkos library to ensure performance portability. Our goal is to provide a user-friendly RMHD code capable of running efficiently on current and future exascale supercomputers.}
   {The radiative transfer module is based on the M1 approximation and implemented using a split explicit–implicit scheme. A reduced speed of light approximation is employed to alleviate the timestep constraint imposed by radiation. The module supports several radiation Riemann solvers and Cartesian, cylindrical, and spherical geometries in one, two, and three dimensions. The implicit step relies on a simple matrix inversion, ensuring both robustness and high performance. Users can choose between built-in opacity models or supply tabulated opacities and custom user-defined functions.}
   {The radiative module of \idefix{} demonstrates excellent performance on accelerated architectures, including the AMD MI250X and MI300 partitions of the AdAstra supercomputer. On MI250X nodes, it achieves up to $7.6\times 10^8$ cell updates per second per node, at a computational cost only 1.6 times higher than a pure magnetohydrodynamic simulation. These results establish \idefix{} as a fast, robust, and portable RMHD code suitable for the wider community.}
   {}
   

   \maketitle
%
\nolinenumbers
\defcitealias{scepi2024}{SBD24}
\defcitealias{Shakura}{SS73}

\section{Introduction}
Radiation plays a central role in the dynamics of many astrophysical systems, often dominating or significantly altering their evolution. Whether generated in situ or irradiating the system from the outside, radiation can often be the dominant force at work, even counteracting gravity in the most extreme cases. This is the case in accreting black holes \citep{hirose2009,jiang2019}, stellar winds \citep{kudritzki2000} and disk winds \citep{proga1998}, galactic disks \citep{thompson2005}, and star-forming regions \citep{norman2010}, to name only a few examples. Yet, even when it is not dynamically dominant, radiation remains an indispensable component of astrophysical fluid dynamics because it regulates the thermodynamics of the flow, thereby determining the relative importance of thermal, magnetic, and gravitational energy. Consequently, radiation must be evolved self-consistently with the fluid dynamics, what is known as radiation magnetohydrodynamics (RMHD). 

Coupling radiative transfer to fluid dynamics, however, requires substantial approximations, as solving the full frequency- and angle-dependent radiative transfer equation is often computationally prohibitive, though it is feasible in some cases using Monte Carlo radiative transfer \citep{miller2019,miller2020,dexter2021}. A common simplification is the gray approximation, in which opacities are averaged over frequency. With the advent of exascale computing, simulations that employ multiple frequency groups have become feasible, although such multigroup radiative transfer calculations remain relatively rare \citep{jiang2025,melon2025,roth2025}. Another widely used simplification is to take angular moments of the radiative transfer equation; this yields the moment equations of radiative transfer, which require a closure relation. The most common closure, flux-limited diffusion (FLD), assumes that the radiative flux is proportional to the gradient of the radiative energy density; while computationally efficient and widely used \citep{hirose2009,flock2017}, this approach loses directional information and is inadequate in many astrophysical regimes. A higher-order closure is the M1 approximation, in which the radiation pressure tensor is expressed as a nonlinear function of the radiative energy and flux, ensuring correct behavior in both the optically thick (Eddington) and optically thin (free-streaming) limits \citep{minerbo1978,levermore1984,ripoll2001,gonzalez2007,rosdahl2013,sadowski2013,skinner2013}. The M1 closure has been successfully applied in a variety of astrophysical contexts, including accretion disks \citep{sadowski2013,mckinney2014,melon2021,liska2022}, galaxy evolution \citep{rosdahl2013,rosdahl2015},  and core collapse supernovae and neutron star mergers, where the radiative fluid is replaced by a neutrino fluid \citep{mezzacappa2020,foucart2023}. While there are higher-order approaches that keep more information about the directionality of the fluid, such as the variable Eddington tensor with a closure made using a Monte Carlo scheme \citep{foucart2018} or a short-characteristic scheme \citep{davis2012}, we chose here to adopt the M1 approximation for its robustness, performance, and ease of implementation.

In this paper, we present the radiation module we implemented in the magnetohydrodynamic  (MHD) public code \idefix{} \citep{lesur2023}. The main advantage of \idefix{} is that it uses the Kokkos library to ensure performance portability across a large variety of architectures, including accelerated architectures, in anticipation of the exascale era. We detail the method that we developed for the radiative transfer module in Sect. \ref{sec:rad_module}, the tests of the method  in Sect. \ref{sec:tests}, and the performance of the radiative version of \idefix{}  in Sect. \ref{sec:perf}, before concluding in Sect. \ref{sec:conclusion}.

\section{Radiative transfer module of \idefix{}}\label{sec:rad_module}

In this section we present the radiative module that we implemented in the \idefix{} code solving for the radiative transfer moment equations in the M1 approximation. The module closely follows the implementation from \cite{melon2019} and \cite{melon2021} except for a few notable differences that we highlight below. 

\subsection{Equations}
The full system for the RMHD equations is
\begin{align}
    &\partial_t \rho + \nabla \cdot \left[ \rho \mathbf{v} \right] = 0,\label{eq:continuity} \\
    &\partial_t (\rho \mathbf{v}) + \nabla \cdot \left[ \rho \mathbf{v} \otimes \mathbf{v} - \mathbf{B} \otimes \mathbf{B} + \mathcal{P} \mathbb{I} \right] = -\rho \nabla \psi + \mathbf{G}, \\
    &\partial_t \mathbf{B} + \nabla \times \left[ - \mathbf{v} \times \mathbf{B} \right] = 0, \\
    &\partial_t E + \nabla \cdot \left[ (E + \mathcal{P}) \mathbf{v} - \mathbf{B}(\mathbf{B} \cdot \mathbf{v}) \right]  \nonumber\\
    &\phantom{\partial_t E + \nabla \cdot } = -\rho \mathbf{v} \cdot \nabla \psi + c G^0 - \nabla \cdot \mathbf{F}_\mathrm{irr}, \label{eq:energy}\\ 
    &\partial_t E_r + \nabla \cdot( \hat{c}\mathbf{F}_r) = -\hat{c}\:G^0,\label{eq:rad1} \\ 
    &\partial_t \mathbf{F}_r + \nabla \cdot (\hat{c} \mathbb{P}_r) = - \hat{c}\:\mathbf{G}, \label{eq:rad2}
\end{align}
where $\rho$ is the density, $\mathbf{v}$ is the velocity, $\mathbf{B}$ is the magnetic field, $\mathcal{P}\equiv P +B^2/2$ is the total pressure with $P$ the thermal pressure, $\mathbb{I}$ is the identity tensor, $\psi$ is the gravitational potential, $\mathbf{F}_\mathrm{irr}$ is the irradiation flux, and $E$ is the total gas energy density: 
\begin{equation}
    E = e + \rho v^2/2 + B^2/2
,\end{equation}
with $e$ the internal energy density.

In the last two equations, $c$ is the speed of light, $\hat{c}$ is the reduced speed of light (see Sect. \ref{sec:reduced_c}), $E_r$ is the radiative energy, $F_r$ the radiative energy flux and $\mathbb{P}_r$ the radiative energy tensor defined, respectively, from the direction and frequency-dependent specific intensity, $I_v(t,\mathbf{x},\mathbf{n})$ as
\begin{align}
    E_r &= \frac{1}{c} \int_0^\infty d\nu \oint d\Omega \, I_\nu(t, \mathbf{x}, \mathbf{n}) \\
    F_r^i &= \frac{1}{c} \int_0^\infty d\nu \oint d\Omega \, I_\nu(t, \mathbf{x}, \mathbf{n})\, n^i \label{eq:def_flux}\\ 
    P_r^{ij} &= \frac{1}{c} \int_0^\infty d\nu \oint d\Omega \, I_\nu(t, \mathbf{x}, \mathbf{n})\, n^i n^j.
\end{align}
Note that Eq. \ref{eq:def_flux} contains a factor, $1/c,$ that is usually not present in the definition of the radiative flux (see \citealt{mihalas2013}). The expression of Eqs. \ref{eq:rad1} and \ref{eq:rad2} takes into account this extra factor.

To close the set of the radiation moment equations, we used the closure from \cite{levermore1984} relating the radiation pressure to the radiation energy and the radiation flux:
\begin{equation}
P^{ij}_r = D^{ij} E_r,
\end{equation}
with
\begin{equation}
D^{ij} = \frac{1 - \xi}{2} \delta^{ij} + \frac{3\xi - 1}{2} n^i n^j,
\end{equation}
and
\begin{equation}
\xi = \frac{3 + 4f^2}{5 + 2\sqrt{4 - 3f^2}},
\end{equation}
where $\delta^{ij}$ is the Kronecker delta symbol, $\mathbf{n}=\mathbf{F_r}/\lVert\mathbf{F_r}\rVert$ is a unit vector pointing in the direction of the radiative flux, and $f$ is the so-called reduced flux defined as
\begin{equation}
f\equiv \frac{\lVert\mathbf{F_r}\rVert}{E_r}\le 1.
\end{equation}
The condition that $f\le1$ must be ensured at all times to keep a physical solution. We discuss this further in Sect. \ref{sec:hyperbolic}.\\

The source terms in the radiation equations are defined as 
\begin{align}
    \tilde{G}^0 &= \kappa_\mathrm{P} \, \rho \left( \tilde{E}_r - a_R T^4 \right) \label{eq:source1} \\
    \tilde{\mathbf{G}} &= \chi \, \rho \, \tilde{\mathbf{F}}_r, \label{eq:source2}  
\end{align}
where the tildes denotes quantities in the fluid rest frame. $\kappa_\mathrm{P}$ is the Planck absorption opacity, $\chi\equiv\kappa_\mathrm{R}+\sigma$ where $\kappa_\mathrm{R}$ and $\sigma$ are the Rosseland absorption and scattering opacities, respectively. It is customary to Lorentz-transform the fluid rest frame source terms in the coordinate frame to find $G^0$ and $\mathbf{G}$. However, since our main application is for protoplanetary disks, where we typically have $\beta\equiv \lVert\mathbf{v}\rVert/c<10^{-3}$, we neglected the Lorentz transform and set $G^0 \equiv \tilde{G}^0$ and $\mathbf{G} \equiv \tilde{\mathbf{G}}$. 

Note that in the nonrelativistic regime of protoplanetary disks, most relativistic corrections arising from the Lorentz transform (typically of order $\mathcal{O}(\beta)$ or $\mathcal{O}(\beta^2)$) can be safely neglected. However, a correction term of order $\mathcal{O}(\tau \beta)$ may still be relevant. This term becomes significant in the dynamic diffusion limit, where the timescales for photon diffusion and advection are comparable. In this regime, photons diffusing through an optically thick medium are advected with the flow.

In this work, we neglected this $\mathcal{O}(\tau \beta)$ correction, as its magnitude is generally comparable to the numerical error of our scheme for $\tau \beta \approx 1$~\citep{skinner2013}. Future studies focusing on accretion disks around compact objects will incorporate all relativistic corrections to address their potential impact in more extreme environments.

\subsection{Units}
The MHD version of \idefix{} is dimensionless. However, radiation naturally introduces units into the system of equation. To keep it conceptually simple for a user, we ask the user to set units for the density, velocity and length. These units are then used to explicitly re-dimension the radiation energy, the radiation flux and the temperature in the source terms. The hyperbolic part of the radiation equation remains dimensionless. 

\subsection{Implementation details}

\idefix{} leverages abstract C++ class templates to enhance modularity, improve performance, and minimize code duplication. The framework employs a single, generic class template to represent all fluid types, with the template parameter defining the specific physics to be solved: whether for classical hydrodynamics, MHD, dust fluids (zero pressure), or radiative fluids. This design eliminates redundancy in critical components such as the Riemann solver, reconstruction scheme, and boundary conditions.
Additionally, the use of class templates enables seamless extensibility to multigroup radiative transfer. This is achieved by instantiating multiple radiative fluid objects within a vector, requiring no structural changes to the existing implementation.

\subsection{Integration scheme}
We use a split operator scheme to integrate the system of equations from \ref{eq:continuity} to \ref{eq:rad2}. Namely, we perform a (magneto)-hydrodynamic step, then we do a radiation transport step and finally add the radiation source terms. This general method is followed for each sub-step of the time integration loop of \idefix{}. For RK2, this is equivalent to the IMEX1 method that is used in \cite{melon2019}. 
We followed \cite{melon2021} by using an explicit integration for the radiation transport (the hyperbolic part of the radiation equations) and an implicit integration for the radiation source terms solving radiation-matter coupling.

For a Euler time integration scheme, this is what a step evolving the radiation field looks like:
\begin{enumerate}
    \item Set the boundary conditions 
    \item Convert the primitive variables to conservative variables
    \item Reconstruct the primitive variables on the faces 
    \item Enforce $f\le 1$ condition on the faces
    \item Compute wave speed at cell interface
    \item Compute Riemann flux at cell interface
    \item Evolve cell-centered conservative variables from flux divergence in each direction
    \item Evolve cell-centered conservative variables from local source terms
    \item Convert conservative to primitive variables
    \item Update time step
\end{enumerate}
We note that in radiation hydrodynamics the sets of primitive variables and conservative variables for the radiation field are identical, namely ($E_r$, $\textbf{F}_r$). Hence, the primitive to conservative (or vice-versa) transformations are trivial.

\subsection{Hyperbolic solver}\label{sec:hyperbolic}

Radiation transport focuses on the left-hand side of Eqs.~\ref{eq:rad1} and~\ref{eq:rad2}, which corresponds to the hyperbolic component of the system. Analogous to the MHD equations, we solved this hyperbolic part using a high-order finite-volume Godunov method. In each computational cell, the primitive variables are first reconstructed at the cell faces using either a slope-limited linear scheme or higher-order methods so that
\begin{align}
E_{r,f} = \mathcal{R}(E_{r,c})\\
F_{r,f}^i = \mathcal{R}(F_{r,c}^i),
\end{align}
where $E_{r,f}$, $E_{r,c}$, $F_{r,f}^i$ and $F_{r,c}^i$ are the radiative energy and the radiative flux in the direction $i$ on the face and cells, respectively, and where $\mathcal{R}()$ denotes the standard reconstruction method of \idefix{}  such as a piecewise linear reconstruction using the van Leer slope limiter (PLM), compact third-order reconstruction (LimO3; \citealt{cada2009}) or piecewise parabolic reconstruction using an extremum-preserving limiter (PPM; \citealt{colella1984}, \citealt{colella2008}). Using these reconstructed variables, a Riemann problem is solved at each interface to compute the inter-cell fluxes of radiation energy and radiation flux in all spatial directions.

In our current implementation, the user choice of reconstruction order and of slope limiter are applied to both the radiation field and hydrodynamic fluid. However, we find that radiation requires a special reconstruction scheme. Indeed, the physical constraint $f\le 1$ can be violated on the (reconstructed) faces even if it is met at the cell centers. In a more general way, the values of $f$ computed on the cell centers and on the cell faces are likely to differ. We find that this discrepancy between cells and faces can lead to unwanted behaviors in the free-streaming limit (see Sects. \ref{sec:riemann} and \ref{sec:free_streaming_beam}). To avoid this issue, \cite{melon2019} switch to a flat reconstruction scheme so that $E_{r,f} = E_{r,c}$ and $F_{r,f}^i = F_{r,c}^i$ whenever $||\mathcal{R}(F_{r,c}^i)||>\mathcal{R}(E_{r,c})$. In the remainder of the paper, we call this reconstruction the ``PLUTO reconstruction scheme."

However, we find that the PLUTO reconstruction scheme tends to render the scheme more diffusive and lead to numerical artifacts (see Sects. \ref{sec:riemann} and \ref{sec:free_streaming_beam}). In order to partially keep the benefit of high-order reconstruction scheme, we propose here an ``$f$-preserving reconstruction scheme." The idea is that, after the reconstruction of $E_r$ and $F_r^i$ on the faces, we check if $||\mathcal{R}(F_{r,c}^i)||>\mathcal{R}(E_{r,c})$ and if $||\mathcal{R}(F_{r,c}^i)||/\mathcal{R}(E_{r,c}) < ||\textbf{F}_{r,c}||/E_{r,c}$. If one of these conditions is met, we recompute the fluxes on the faces in each direction with the constraint that the value of $f$ must be equal on the faces and at the center of the cell so that
\begin{align}
&E_{r,f} = \mathcal{R}(E_{r,c}),\\
&F_{r,f}^i = \mathcal{R}(F_{r,c}^i)\:\:\:\:\:\:\:\mathrm{if}\:\frac{||\mathcal{R}(F_{r,c}^i)||}{\mathcal{R}(E_{r,c})}\le1\:\:\mathrm{and}\:\:\frac{||\mathcal{R}(F_{r,c}^i)||}{\mathcal{R}(E_{r,c})} > \frac{||\textbf{F}_{r,c}||}{E_{r,c}},\\\label{eq:f_reconstruct}
&F_{r,f}^i = F_{r,c}^i \frac{\mathcal{R}(E_{r,c})}{E_{r,c}}\:\:\:\:\:\mathrm{otherwise.}
\end{align}
The second condition in Eq. \ref{eq:f_reconstruct} is introduced to prevent what we refer to as "beam widening" in the free-streaming regime. Without this constraint, the reconstruction scheme may artificially introduce a transverse pressure-like component into an otherwise free-streaming beam, thereby causing an unphysical broadening of the beam (see Sect. \ref{sec:free_streaming_beam} for further details).

We implemented three Riemann solvers for radiation, Lax-Friedrichs-Rusanov, HLL \citep{gonzalez2007}, and HLLC \cite{melon2019}. In practice, for our applications we only use HLL but all three solvers have been tested on all test cases. We do not describe here the details of the solvers and refer the reader to \cite{melon2019} for details.

 \subsection{Signal speed}

We computed the wave velocities as in \cite{skinner2013}. The M1 system of equations has four eigenvalues, two of which are degenerate; they are
\begin{equation}
\lambda_{1,3} = \frac{f \cos(\theta)\pm\zeta(f,\theta)}{\sqrt{4-3f^2}}
\end{equation}
\begin{equation}
\lambda_2 = \frac{3\xi(f)-1}{2f}\cos(\theta),
\end{equation}
where 
\begin{align}
\zeta(f,\theta)
&=
\left[
\frac{2}{3}\left(4 - 3f^2 - \sqrt{4 - 3f^2}\right)
\right. \notag\\
&\qquad \left.
+ 2\cos^2(\theta)
\left(2 - f^2 - \sqrt{4 - 3f^2}\right)
\right]^{1/2}.
\end{align}
and $\theta$ is the angle between $\textbf{F}_r$ and the direction in which we are evaluating the Riemann flux.

In the free-streaming-limit ($f\to1$), we recover that all eigenvalues $\lambda_{1,2,3}\to \cos(\theta)$ so that there are equal to $\pm1$ along the direction of the radiative flux and zero in the direction perpendicular to it. In the diffusion limit ($f \to 0$), we recover $\lambda_{1,3}\to \mp 1/\sqrt{3}$ and $\lambda_2\to0$ in all directions, in the Eddington limit. In our scheme, only $\lambda_{1,3}$ are computed. However, $\lambda_2$ is useful for understanding the structure of the Riemann problem tests that we present in Sect. \ref{sec:riemann}. It is the propagation velocity of the equivalent of a contact discontinuity in hydrodynamics (meaning it is canceled out in the discontinuity comoving frame). 

In the very optically thick limit, the diffusion timescale can be much larger than $\lambda_{1,3}$ leading to excessive numerical diffusion. To circumvent this issue, we adopted the method used in \cite{sadowski2013}, \cite{rosdahl2015}, and \cite{melon2019} and corrected the signal speed as follows:
\begin{align}
 \lambda_{L} &= \max(\lambda_L,-\frac{4 \hat{c}}{3\tau_\mathrm{cell}}) \\
 \lambda_R &= \min(\lambda_R,\frac{4 \hat{c}}{3\tau_\mathrm{cell}}),
\end{align}
where $\lambda_L$ and $\lambda_R$ are respectively the maximum and minimum signal velocities across the left and right faces, and $\tau_\mathrm{cell}=\rho\chi\Delta x$ is the optical depth within a cell of maximum size $\Delta x$.

\subsection{Reduced speed of light}\label{sec:reduced_c}

The use of an explicit integration scheme has the drawback of drastically reducing the timestep when the characteristic MHD wave speeds are much slower than the speed of light. To alleviate this issue, we used the reduced speed of light approximation \citep{Gnedin2001}, where $\hat{c}$ is an input parameter taken to be constant across the domain. A direct consequence of the reduced speed of light approximation is the violation of the usual form of the energy conservation \citep{melon2021}. Without gravity or dissipative source terms, the rescaled quantities that are effectively conserved are
\begin{gather}
E_\mathrm{tot} = E + (c/\hat{c})E_r,\label{eq:conservation_energy}\\
\mathbf{m}_\mathrm{tot} = \rho\mathbf{v} + (1/\hat{c})\mathbf{F}_r.
\end{gather}
The reduced speed of light approximation does guarantee the convergence to the correct steady-state solution but can lead to artifacts in the temporal evolution leading to steady-state. To stay as close as possible to the real solution, one needs to keep the timescale ordering, i.e., the reduced speed of light should still exceed all other characteristic speeds of the system and the diffusion time should be shorter than the dynamical time (see Sect. \ref{sec:shocks} for more details).

\subsection{Radiation-matter coupling terms}\label{sec:implicit}

Radiation matter-coupling deals with the right hand-side of Eqs. \ref{eq:rad1} and \ref{eq:rad2}. As stated above, it is customary to Lorentz transform the source terms from the fluid-frame to the laboratory-frame so that the source terms usually involve both hydrodynamic and radiation-related variables. However, since our main application is for protoplanetary disks, we decided to neglect the Lorentz transform as $\beta \ll 1$. This means that the source terms only depend on the density and temperature for the hydrodynamic variables and not on the velocity anymore. 

To solve Eqs. \ref{eq:rad1} and \ref{eq:rad2}, we used an implicit method. Using Eqs. \ref{eq:source1} and \ref{eq:source2} we can discretize and rewrite Eqs. \ref{eq:rad1} and \ref{eq:rad2} (omitting the hyperbolic part of the equation) as 
\begin{align}
&E_r^{n+1}\left( 1+ \Delta t/t_\mathrm{abs,red} \right) -\Delta t/t_\mathrm{abs,red} a_R \left({T^{n+1}}\right)^4 = E_r^{n} \label{eq:source_discrete1},\\
&F_r^{n+1}\left(1+\Delta t /t_\mathrm{scatt,red}\right) = F_r^{n},\label{eq:source_discrete2}
\end{align}
where $n$ denotes the integration step, $t_\mathrm{abs,red}\equiv 1/(\hat{c} \kappa_\mathrm{P}\rho^n)$ is the characteristic timescale for radiation absorption associated with the reduced speed of light and $t_\mathrm{scatt,red}\equiv 1/(\hat{c} \chi\rho^n)$ is the typical timescale for radiation scattering associated with the reduced speed of light.

Equation \ref{eq:source_discrete2} is trivially solved, but Eq. \ref{eq:source_discrete1} requires an equation on $T^{n+1}$ to close the system. To this end, we use the conservation of internal energy, 
\begin{equation}
C_v\partial_t T = c\kappa_\mathrm{P} \, \rho \left( E_r - a_R T^4 \right),\label{eq:internal}
\end{equation}
which we discretize as 
\begin{equation}
    C_vT^{n+1}+\Delta t/t_\mathrm{abs}a_R(T^{n+1})^4 - \Delta t/t_\mathrm{abs} E_r^{n+1} = C_vT^n,
\end{equation}
where $t_\mathrm{abs}\equiv 1/(c \kappa_\mathrm{P}\rho^n)$ and $C_v$ is the calorific capacity.

We then followed \cite{hayes2003} and \cite{commercon2011} and assumed that $T$ does not change significantly over one time step so that we can write $({T^{n+1})}^4 = 4T^{n+1}({T^n})^3-3({T^{n}})^4$. We can then rewrite Eqs. \ref{eq:source_discrete1} and \ref{eq:internal} in the following way:

\begin{equation}\label{eq:implicit_matrix}
\begin{bmatrix}
1+\gamma_\mathrm{abs,red} & -4\gamma_\mathrm{abs,red}a_R{T^n}^3\\
-\gamma_\mathrm{abs} & C_v+4\gamma_\mathrm{abs}a_R{T^n}^3
\end{bmatrix}
\begin{bmatrix}
E_r^{n+1}\\
T^{n+1}
\end{bmatrix} = \begin{bmatrix}
E_r^n-3\gamma_\mathrm{abs,red}a_R{T^n}^4\\
C_vT^n+3\gamma_\mathrm{abs}a_R{T^n}^4
\end{bmatrix}\\
,\end{equation}
where $\gamma_\mathrm{abs,red}\equiv\Delta t/t_\mathrm{abs,red}$ and $\gamma_\mathrm{abs}\equiv\Delta t/t_\mathrm{abs}$.
To find $E_r^{n+1}$, we inverted this $2^2$ matrix for each cell, which we could do analytically. Finally, to complete the implicit step, we computed the new gas energy and momentum in each direction by using the rescaled conservation of energy, Eq. \ref{eq:conservation_energy}. While the implicit part of our scheme is currently written for single group radiative transfer, our implementation can trivially be extended to multigroup by solving for a $(m+1)^2$ matrix instead of a $2^2$, where $m$ is the number of frequency groups (see Sect. \ref{sec:implicit}). 

We preferred to use this implicit solver rather than the fixed point method used in \cite{melon2021} because we find that it is more robust (less prone to overshooting or undershooting the solution) and performs better since it is not an iterative method. For completeness, we also implemented the fixed-point method described in \cite{melon2019} but refer the reader to the aforementioned paper for details as we did not use this method in the tests presented here.

\subsection{Irradiation flux}

To treat irradiation by an external object, we allow the user to define an irradiation flux, the divergence of which acts as a source term in the energy conservation equation, Eq. \ref{eq:energy}. This irradiation flux is not part of the radiation scheme; it is defined through a user function that updates instantaneously the irradiation flux throughout the entire domain at each time step. We find that the best way to implement this term is to incorporate it to the implicit scheme. To do so, we rewrite Eq. \ref{eq:conservation_energy} to take into account the energy deposited by the irradiation flux so that 
\begin{equation}
    E_\mathrm{tot} = E + (c/\hat{c})E_r -\nabla \cdot \mathbf{F}_\mathrm{irr}\times \Delta t,\label{eq:conservation_energy_irr}
\end{equation}
where $\Delta t$ is the timestep. We then also rewrite the bottom term of the right hand side of Eq. \ref{eq:implicit_matrix}, $S_1$, to 
\begin{equation}
S_1 =  C_vT^n+3\gamma_\mathrm{abs}a_R{T^n}^4 - \nabla \cdot \mathbf{F}_\mathrm{irr}\times \Delta t,
\end{equation}
solve for this new system to find $E_r^{n+1}$, and finally use Eq. \ref{eq:conservation_energy_irr} to find $E^{n+1}$. We find that adding the irradiation source terms in the implicit step provides a much better accuracy compared to a split-scheme, where irradiation would be added afterward, and prevents an overshooting of the solution in the test case (see Sect. \ref{sec:irradiation_test}).

\section{Radiation test problems}\label{sec:tests}

In the first two tests, the optically thin "shock tubes" and the free-streaming beam, the choice of $\hat{c}$ does not impact the results as we test only the hyperbolic part of Eqs. \ref{eq:rad1} and \ref{eq:rad2}. In the third, fourth, and seventh tests, the shadow tests, the pulse tests, and the stellar irradiation tests, the hydrodynamic fluid is static, so the  Courant–Friedrichs–Lewy  condition is only determined by the radiation subsystem. In such cases, the choice of $\hat{c}$ is also without consequences. Hence, the only tests where the choice of $\hat{c}$ matters are the fifth and sixth tests, the radiative shocks and vertical diffusion tests.

\subsection{Optically thin "shock tube" tests}\label{sec:riemann}

\begin{figure}
    \centering
    \includegraphics[width=90mm]{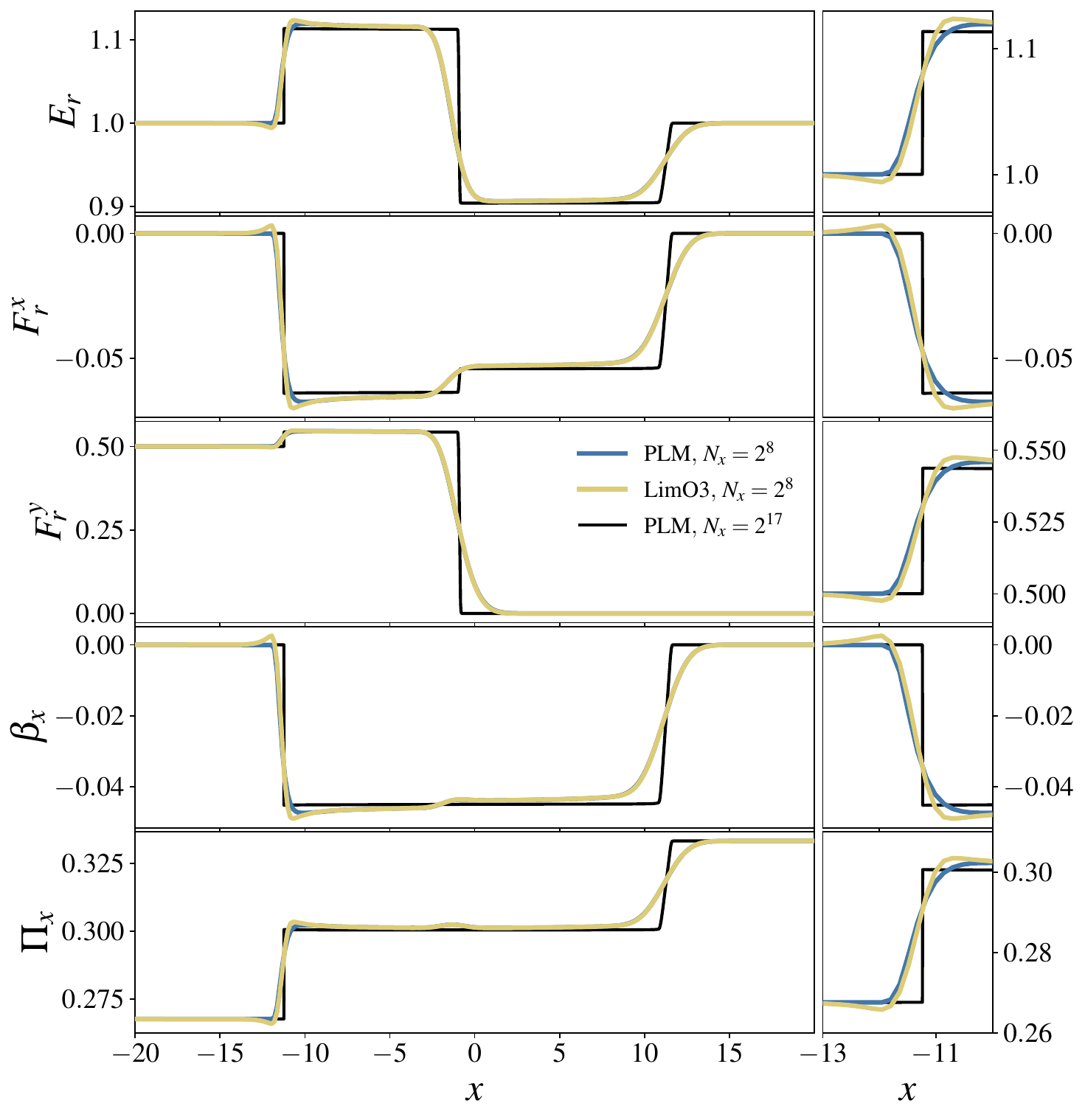}
    \caption{First test of an optically thin shock tube. The blue and green lines show the results of simulations with $2^8$ radial cells using a PLM reconstruction and LimO3 reconstruction, respectively. The black line shows the reference solution using PLM reconstruction with $2^{17}$ radial cells. \textit{Left panels}: Full solution. \textit{Right panels}: Zoomed-in view of the left-facing shock at $x\approx11.2$.}
    \label{fig:opt_riemann1}
\end{figure}

\begin{figure}
    \centering
    \includegraphics[width=90mm]{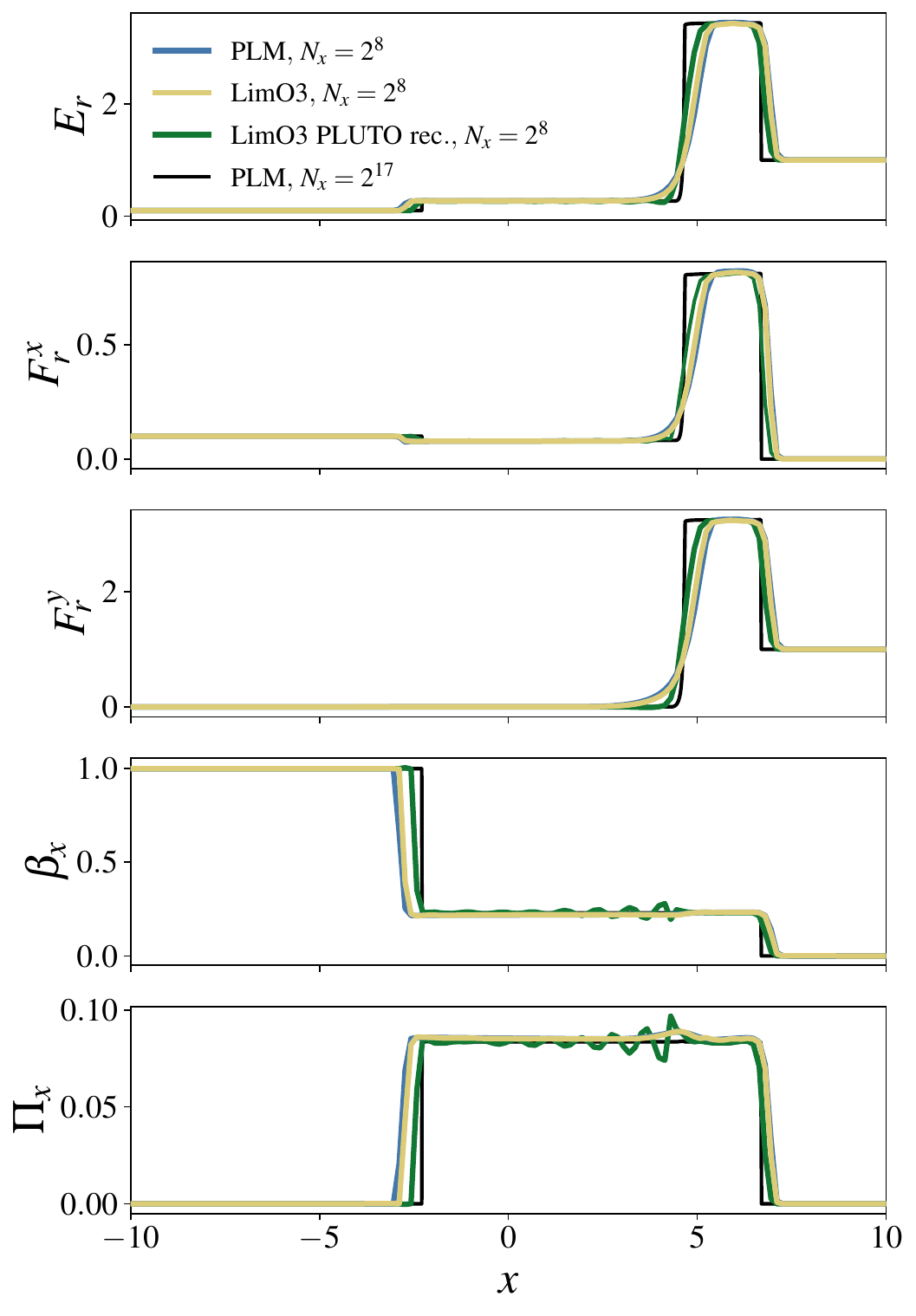}
    \caption{Second test of an optically thin shock tube. The blue and green lines show the results of simulations with $2^8$ radial cells using a PLM reconstruction and a LimO3 reconstruction, respectively, with our reconstruction of $f$ on the faces. The red line shows the result of a simulation with $2^8$ radial cells using a PLM reconstruction with PLUTO's reconstruction of $f$ on the faces. The black line shows the reference solution using PLM reconstruction with $2^{17}$ radial cells.}
    \label{fig:opt_riemann2}
\end{figure}

To test the implementation of the hyperbolic solver, we performed two tests of 1D radiative shock tubes in Cartesian coordinates following \cite{melon2019}. The two tests follow the evolution of two constant states $L$ and $R$, located respectively to the left and right of $x=0$. We neglected any interaction with matter by setting all opacities to zero. The simulation box extends from $x\in[-20,20]$.

The first test is a discontinuity along the $x$-direction in the tangential radiative flux. Specifically, we set as $E_{r,L}=E_{r,R}=1$, $F_{r,x,L}=F_{r,x,R}=0$ and $F_{r,y,L}=1/2$ and $F_{r,y,R}=0$. The radiation field has $f=1/2$ so that it is neither in the free-streaming limit nor in the diffusion limit. We show in Fig. \ref{fig:opt_riemann1} the solution obtained at $t=20$ for the HLL solver with two different reconstruction schemes, the $f$-preserving PLM (blue line) and $f$-preserving LimO3 (green line), at a resolution of $2^8$. We also show the result of a simulation with HLL and a $f$-preserving PLM reconstruction with a resolution of $2^{17}$ for reference (black line). We see in the top three panels the development of a three-wave pattern, with (1) a left-facing shock on the left, (2) a contact-like discontinuity in the middle, and (3) a rightward expansion wave on the right. The two bottom panels show that, at the highest resolution, the fields $\beta_x\equiv(3\xi-1)F_{r,x}E_r/2||\textbf{F}_r||$ and $\Pi\equiv(1-\xi)E_r/2$, which respectively are analogs of the hydrodynamic velocity and pressure, are constant across the contact-like discontinuity. In this test, we see that in contrast to PLM, LimO3 tends to overshoot and undershoot the solution at the discontinuity as can be seen from the right panels that are zoomed on $x\approx 11.2$. We find that this behavior is quite general to all tests and can lead to negative radiative energy in situations where $E_r$ is very small on one side of the discontinuity. In such cases, we needed to manually enforce the positivity of the radiative energy limiting the usefulness of higher-order reconstruction schemes.

The second test is that of a shock in the $x$ direction. Specifically, we set $E_{r,L}=1/10$, $E_{r,R}=1$, $F_{r,x,L}=1/10$ $F_{r,x,R}=0$, $F_{r,y,L}=0,$ and $F_{r,y,R}=1$.  The radiative fluid has $f=1$ on both sides so that we test the ability of the solver to handle the free-streaming limit. Again, we show in Fig. \ref{fig:opt_riemann2} the solution at $t=20$ using the HLL solver but with different reconstruction schemes, PLM and LimO3 (both $f$-preserving), at a resolution of $2^8$. We also show the result of a simulation with a PLM $f$-preserving reconstruction with a resolution of $2^{17}$ and that of a simulation with a LimO3 with the PLUTO reconstruction scheme for reference (see Sect. \ref{sec:hyperbolic}). Again, we see in the top three panels the development of a three-wave pattern, with 1) a left-facing shock on the left, 2) a contact-like discontinuity in the middle and 3) a right-going shock on the right. This test illustrates the need for a careful treatment of the reconstruction on the faces in the free-streaming limit. Indeed, we see in the two bottom panels that our solution with the reconstruction scheme of PLUTO produces spurious oscillations on the left side of the contact-like discontinuity when using LimO3. We checked that this behavior is also observed using the PLUTO code. Our reconstruction scheme does not produce such spurious oscillations; however, it does produce a small peak in $\Pi$ on the left side of the contact-like discontinuity with both PLM and LimO3. Despite this slight loss of accuracy with PLM compared to the reconstruction scheme of PLUTO, we chose to adopt our reconstruction scheme given its improvement in the solution with LimO3.

\begin{figure}
    \centering
    \includegraphics[width=80mm]{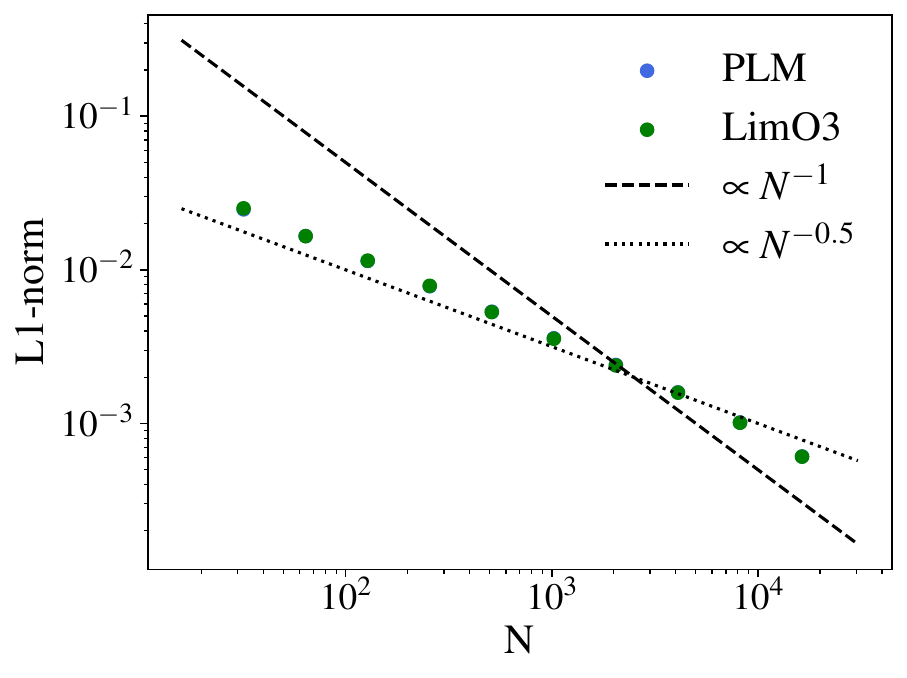}
    \includegraphics[width=80mm]{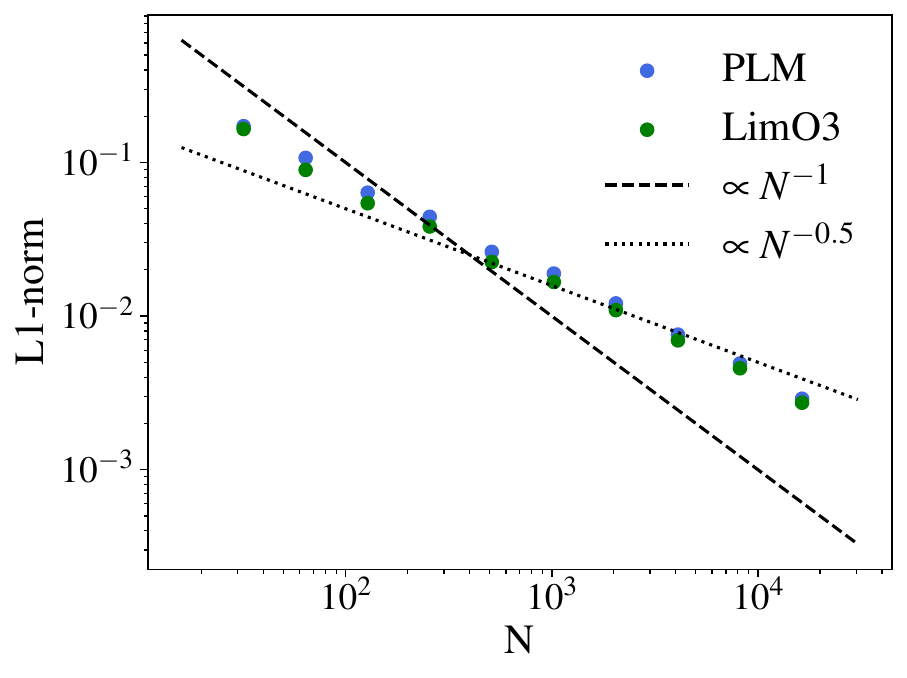}
    \caption{L1 norm error as a function of the number of cells for the first optically thin Riemann problem (\textit{top}) and the second optically thin Riemann problem (\textit{bottom}) computed from a reference solution with $2^{17}$ radial cells. Blue and green points show simulations with PLM and LimO3 reconstructions, respectively.}
    \label{fig:L1_riemann}
\end{figure}

In Fig. \ref{fig:L1_riemann} we show the result of a resolution study for both Riemann tests using second-order PLM and third-order LimO3 reconstructions schemes spanning resolutions going from $2^5$ to $2^14$. We plot the L1-norm error of $E_r$ computed compared to the solution of reference with a resolution of $2^{17}$. We use a Courant number of 0.4 and a RK2 time integrator. We see that both tests scale as $N^{-1/2}$, which is expected for shock solutions \citep{leveque2002}.  

\subsection{Free streaming beam}\label{sec:free_streaming_beam}
To test the multidimensional hyperbolic transport of our radiative scheme, we implement an oblique free streaming radiation beam test \citep{richling2001,gonzalez2007}. We set all opacities to zero so that there is no interaction between radiation and matter. The domain consists of a square Cartesian grid of size $L=5$ in code units. We initialize the background radiation field as $E_r=10^4$ and $F_{r,\:x}=F_{r,\:y}=0$ in code units. The beam is initialized between $x\in[0.5,0.6]$ and $y\in[0.3,0.44]$ with $E_r=10^{12}$ and $F_{r,\:x}=F_{r,\:y}=E_r/\sqrt{2}$. The boundary conditions are outflow everywhere. Note that we do not inject the beam from the boundary condition as in \cite{melon2019} to avoid the implementation of a special boundary condition for oblique injection. Finally, we use a resolution of $300\times300$, a Courant factor of 0.4, the HLL scheme and different reconstruction schemes. 

\begin{figure}
    \centering
    \includegraphics[width=80mm]{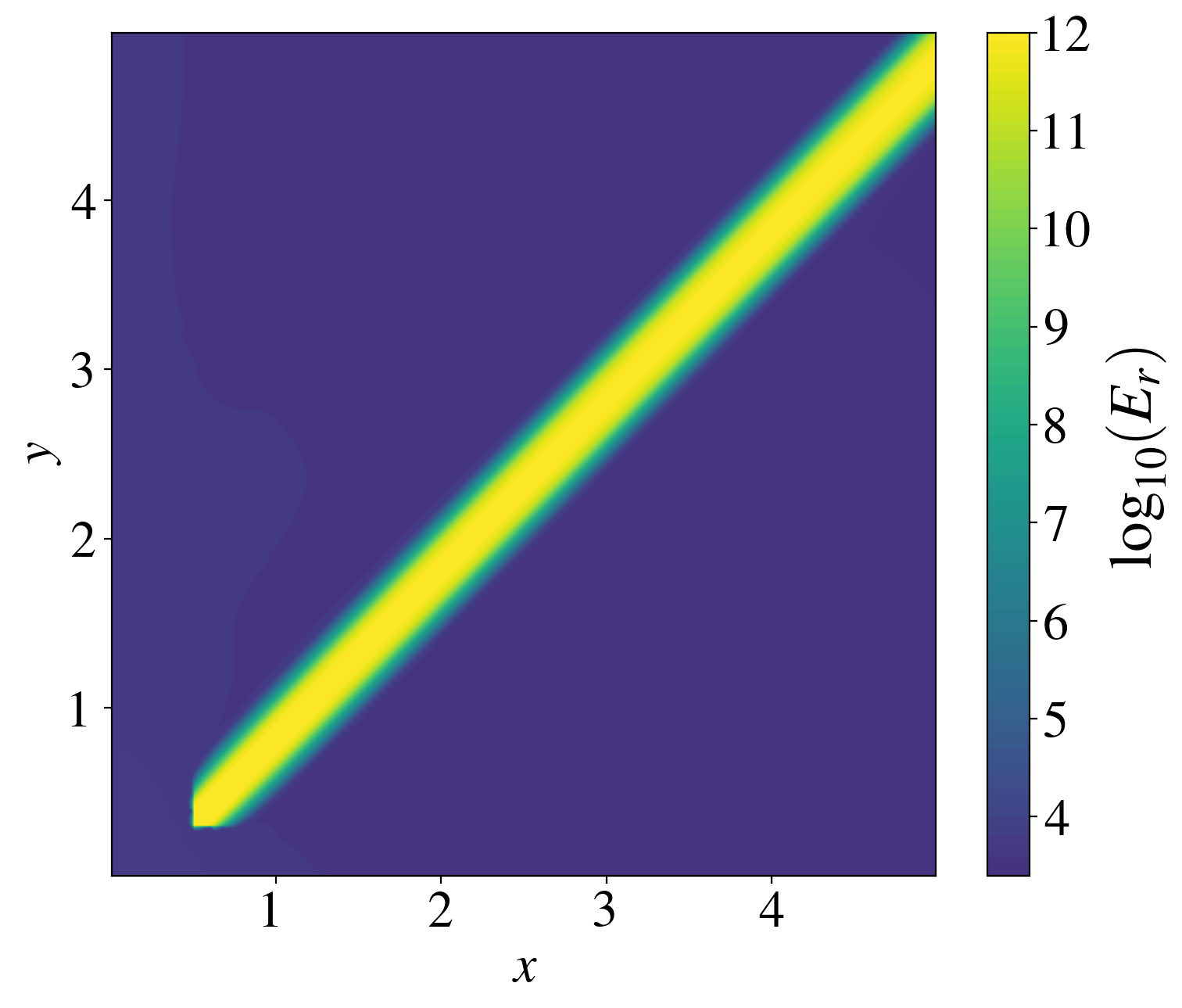}
    \includegraphics[width=80mm]{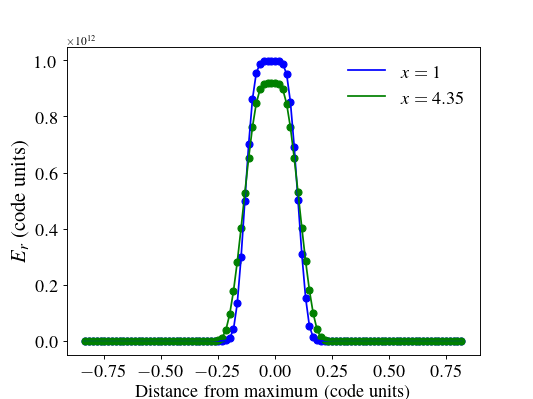}
    \caption{\textit{Top}: Color map of the radiation energy density for the free streaming beam with a resolution of $300\times300$ with a LimO3 $f$-preserving reconstruction scheme. \textit{Bottom}: Vertical cut of the radiation energy density at $x=1$ and $x=4.35$. }
    \label{fig:beam}
\end{figure}

\begin{figure}
    \centering
    \includegraphics[width=80mm]{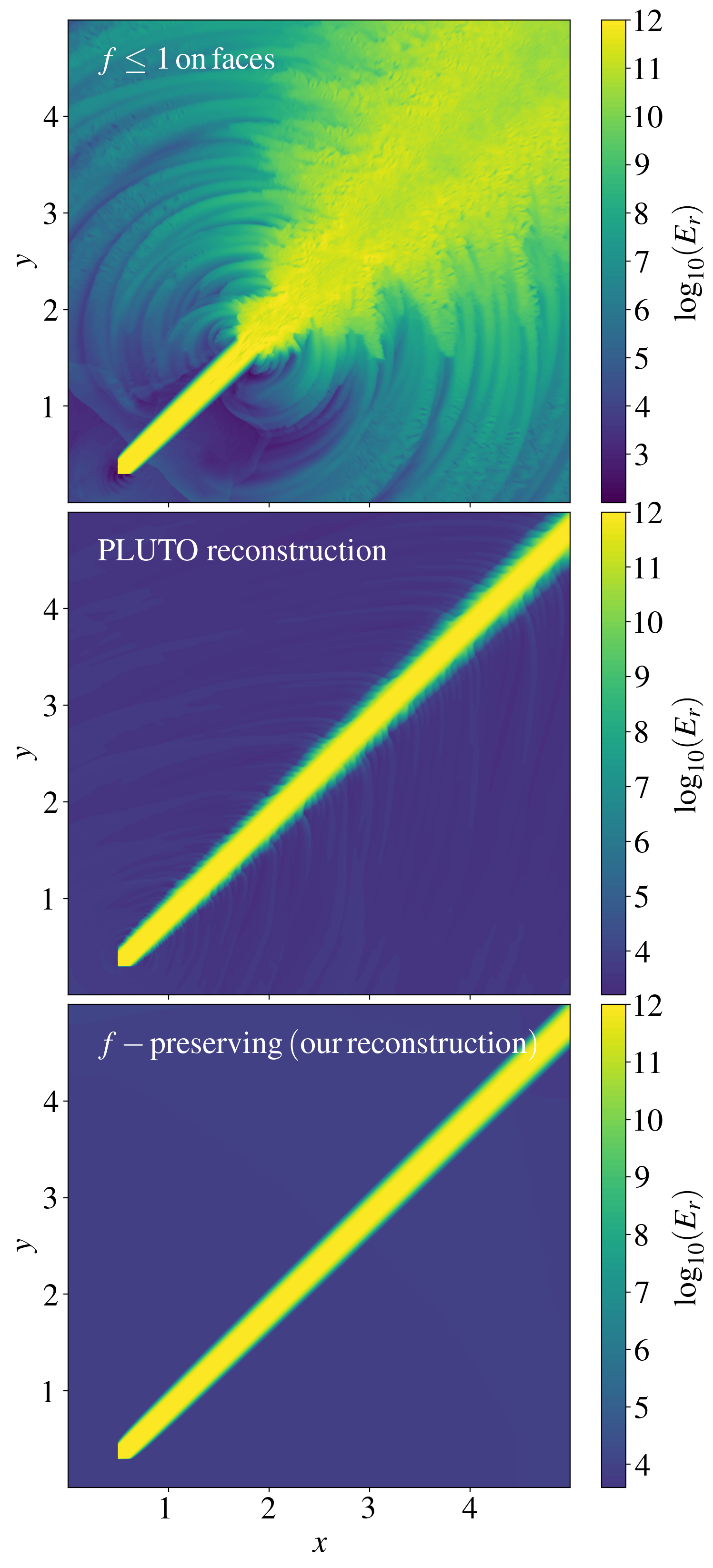}
    \caption{Color maps of the radiation energy density for the free streaming beam test run using different reconstruction scheme. \textit{Top}: Reconstruction scheme where we only impose $f\le1$ on the faces after reconstruction. \textit{Middle}: PLUTO radiative reconstruction scheme. \textit{Bottom}: Our $f$-preserving reconstruction scheme. }
    \label{fig:failure_beam}
\end{figure}

We show in the first panel of Fig. \ref{fig:beam} a color plot of this test using the PLM $f$-preserving reconstruction. As expected, the beam is propagating in a straight line and reaches the outer boundary. As it is propagating the beam widens due to numerical diffusion. To quantify this diffusion, we plot, in the bottom panel of Fig. \ref{fig:beam}, horizontal slices of the beam at $x=1$ and $x=4.35$ that are centered on the maximum value of the profile. We see that the beam maximum decreased by $8\%$ and the beam's full width at half maximum increased by only $7\%$ during its propagation. This is much lower than what is reported in \cite{melon2019} where increase by as much as $40\%$ although we do not use exactly the same metric.

In Fig. \ref{fig:failure_beam} we also illustrate the importance of the choice of reconstruction scheme with a high-resolution test of the free streaming beam with a resolution of $1000\times 1000$. All panels use the LimO3 flux-reconstruction but enforce the physical condition $f\le 1$ in different ways. The first panel shows a reconstruction scheme where we simply enforce that $f\le 1$ on cell-faces. More precisely, we set
\begin{align}
&E_{r,f} = \mathcal{R}(E_{r,c}),\\
&F_{r,f}^i = \mathcal{R}(F_{r,c}^i)\:\:\:\:\:\:\:\mathrm{if}\:\frac{||\mathcal{R}(F_{r,c}^i)||}{\mathcal{R}(E_{r,c})}\le1,\\
&F_{r,f}^i = \mathcal{R}(F_{r,c}^i) \frac{\mathcal{R}(E_{r,c})}{||\mathcal{R}(F_{r,c}^i)||}\:\:\:\:\:\mathrm{otherwise.}
\end{align}
We see that with this simple reconstruction scheme the beam "explodes." We believe that this is due to the accumulation of small deviation between $f$ on the center and faces of the cells that introduce deviation from the free-streaming limit as discussed in Sect. \ref{sec:hyperbolic}. The middle panels show the results of the free streaming beam test using the "PLUTO reconstruction" scheme and our $f$-preserving reconstruction scheme. We see that both of these methods fix the beam explosion and give similar results although the PLUTO reconstruction produces spurious features around the beam that are absent with our reconstruction. 

\subsection{Shadow}
The M1 method offers a significant advantage over the FLD approach by retaining directional information about the radiation field, thereby enabling an accurate representation of shadowing effects. To validate this capability, we conducted the benchmark test proposed in \cite{hayes2003}, \cite{gonzalez2007}, and \cite{melon2019}, which consists of illuminating an opaque obstacle from one side and examining the resulting shadow formed in the downstream region. The setup consists of a rectangular box of dimension $L\times l= 1\:\mathrm{cm}\times 0.3\:\mathrm{cm}$ with a resolution of $140\times40$. The radiation beam is injected from the left boundary along the x-direction with a radiation temperature of $T_r=1740\:K$ and a flux along the x-direction $F_{r,\:x}=E_r$ corresponding to the free-streaming limit. The right hand side horizontal boundary and the top vertical boundary are outflowing while the bottom vertical boundary, at $y=0$, is reflective. The opaque surface is a spheroid of density $\rho_1=10^{3}\:\mathrm{g\:cm^{-3}}$ surrounded by a background with a density of $\rho_0=1\:\mathrm{g\:cm^{-3}}$. To allow for a smooth transition between $\rho_0$ and $\rho_1$, we set 
\begin{equation}
\rho = \rho_0 + \frac{\rho_1-\rho_0}{1+e^\Delta}
,\end{equation}
where 
\begin{equation}
\Delta = 10\left[\left(\frac{x}{x_0}\right)^2+\left(\frac{y}{y_0}\right)^2 -1\right], 
\end{equation}
with $(x_0,y_0)=(0.1,0.06)$. The entire domain is set in thermal equilibrium with $T=T_r=290\:K$. Additionally, the radiative fluxes and gas velocities are initially set to zero. Finally, we set the absorption Rosseland opacity using Kramers' law so that $\kappa=0.1(\rho/\rho_0)(T/T_0)^{-3.5}\:\mathrm{cm^2\:g^{-1}}$. With this choice of opacity, we ensure that the spheroid is optically thick to radiation while the rest of the domain is optically thin. 

\begin{figure}
    \centering
    \includegraphics[width=90mm]{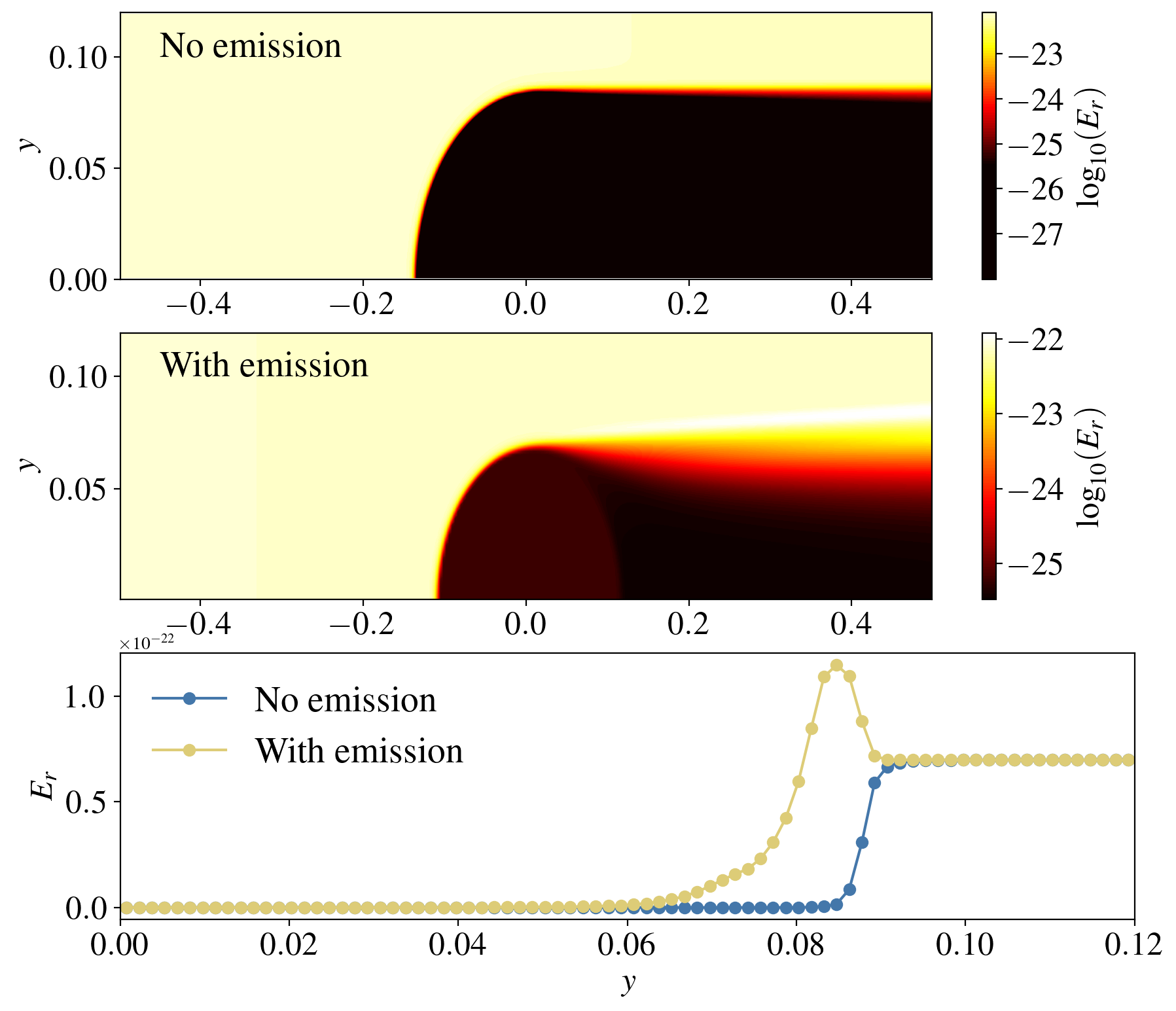}
    \caption{Color map of the radiation energy density without emission source terms (\textit{top}) and with emission source terms (\textit{middle}) for the shadow test run. \textit{Bottom}: Vertical slice of the radiation energy density at $x=0.5$ for the run with emission (yellow line and dots) and without emission (blue line and dots). }
    \label{fig:shadow}
\end{figure}

We show in Fig. \ref{fig:shadow} two color maps of the radiation energy after 10 light crossing time, by which time the beam has long passed the opaque spheroid, reached the rightmost boundary and settled to a steady state. In the top panel, we follow \cite{skinner2013} and remove the source terms responsible for emission of radiative energy by the gas so that radiation can only be absorbed. In practice, this means that the radiation-matter interaction reduces to
\begin{equation}
E_r^{n+1} = \frac{E_r^n}{1+\gamma_\mathrm{abs,red}}
\end{equation}
for the radiative energy instead of Eq. \ref{eq:implicit_matrix}. The top panel of Fig. \ref{fig:shadow} confirms the expected formation of a sharp shadow behind the spheroid, with the transition between the shadowed and illuminated regions confined to a single cell. In the middle panel, we present the case where both emission and absorption terms are included so that we are solving for Eq. \ref{eq:implicit_matrix}. Here, the shadow appears more diffuse, and an overshoot in radiation energy is observed at the transition between the shadow and the beam. Additionally, the radiation energy within the shadow is higher when emission terms are accounted for, as the gas located behind the spheroid contributes to the local light emission.

We attribute the observed over-density in radiation energy to the interaction between the incident beam and the emitted light originating from behind the shadowed region. Overall, our results are quantitatively consistent with those reported in \cite{gonzalez2007}, \cite{skinner2013}, and \cite{melon2019}.

\subsection{Pulses}\label{sec:pulse}

We next tested the ability of our scheme to handle transport of radiation in the optically thin limit (but with matter-light interactions on) in Cartesian and spherical coordinates by performing a pulse propagation test. We initialized a spherically symmetric radiation energy distribution as $E_r=4\pi B(T_r)$, where $B(T)$ is the blackbody function and 
\begin{equation}
    T_r=T_0(1+100 e^{-r^2/w^2})
,\end{equation}
where $r$ is the spherical radius, $T_0=1,05\times10^{-7}$ K, and $w=5$ in code units. The gas is initially at a temperature of $T_0$ so that it is in thermal equilibrium with the radiation away from the pulse. We set $\mathbf{F}_r=\vec{0}$, $\rho=1$ in code units, $\kappa=0$, $\Gamma=5/3$, $C_a=0.4$. Our units are $\mathrm{unit}_v=c$,  $\mathrm{unit}_\rho=1.67\times 10^{-24}\:\mathrm{g\:cm^{-3}}$ and $\mathrm{unit}_L=1.496\times 10^{13}\:\mathrm{cm}$.

\begin{figure}
    \centering
    \includegraphics[width=90mm]{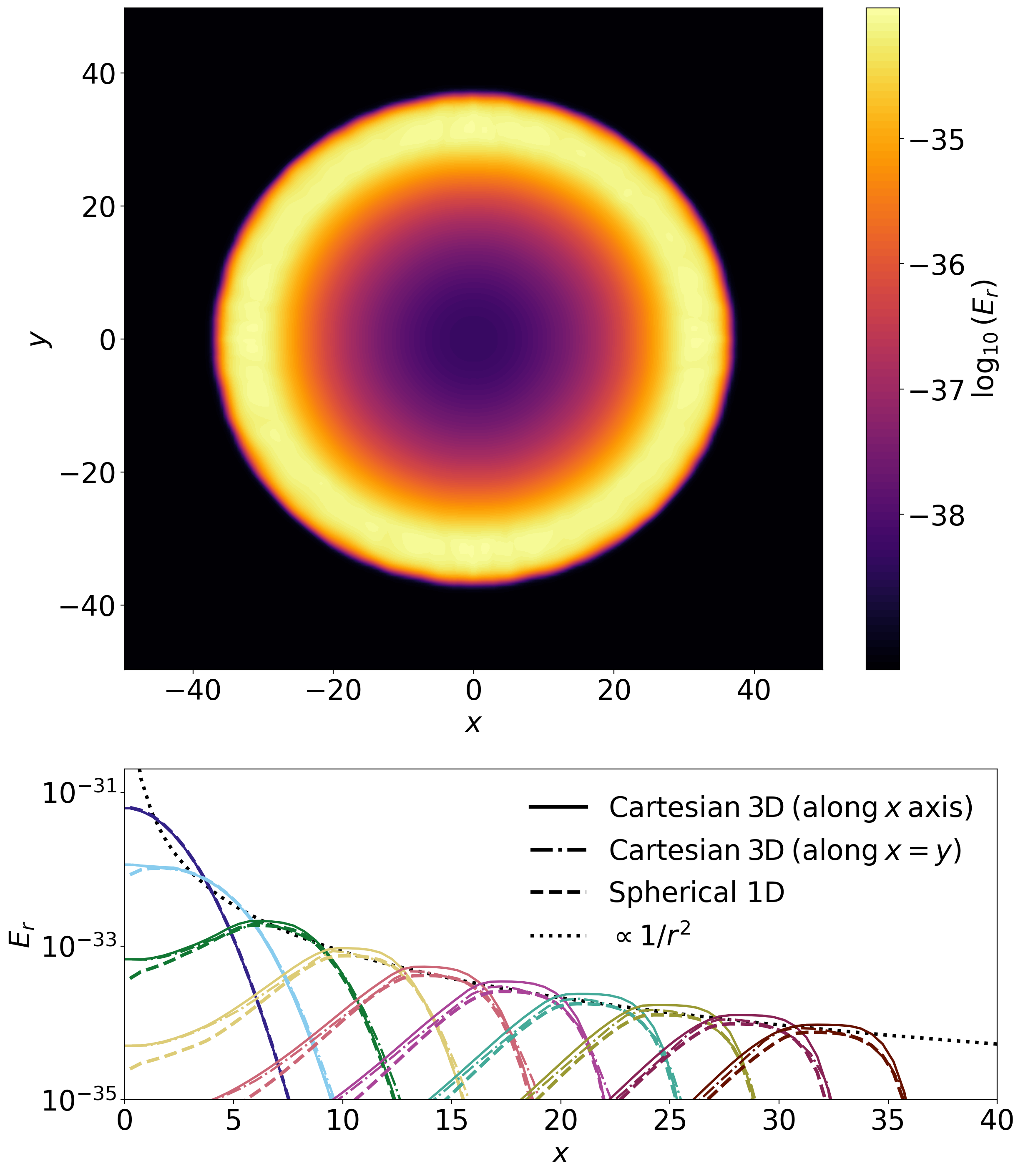}
    \caption{\textit{Top}: Color map of the radiation energy density for the optically thin 3D Cartesian pulse test. \textit{Bottom}: Slices along the $x$ axis and $x=y$ axis for the 3D Cartesian pulse test (solid and dash-dotted lines, respectively) at different times compared to the 1D spherical case (dashed lines). The dotted black line shows the expected decrease in the energy density as $1/r^2$.}
    \label{fig:pulse}
\end{figure}

To ensure an optically thin domain, we set a constant scattering opacity of $\sigma=3.9\times 10^4\:\mathrm{cm^2\:g^{-1}}$. The top panel of Fig. \ref{fig:pulse} shows the energy density of the optically thin pulse at $t=35$ in code units when propagating on 3D on a Cartesian grid of $200\times200\times200$ with $(x,y,z)\in[-50,50]\times[-50,50]\times[-50,50]$. We see that the pulse is spherically expanding as expected. For this resolution, we do not see, with the naked eye, artifacts along each axis due to the Cartesian grid as can be seen in \cite{melon2019}. In the bottom panel, we plot two 1D profiles of the radiation energy density for the 3D Cartesian simulation, one along the $x$ axis at $y=0$ and one along the $x=y$ axis. We also plot the radiation energy density from a 1D spherical simulation for comparison. We see that the spherical simulation decays as $1/r^2$ as expected. The profile of the pulse along the $x=y$ axis in the Cartesian run is almost indistinguishable from the spherical profile. However, along the $x$-axis of the Cartesian simulation the pulse has a larger amplitude than the spherical case. This is an artifact due to propagation along the grid. For consistency, we checked that the total energy of the pulse is the same in the two simulations and that in each case it is conserved to machine precision.  

\subsection{Shocks}\label{sec:shocks}

\begin{figure}
    \centering
    \includegraphics[width=90mm]{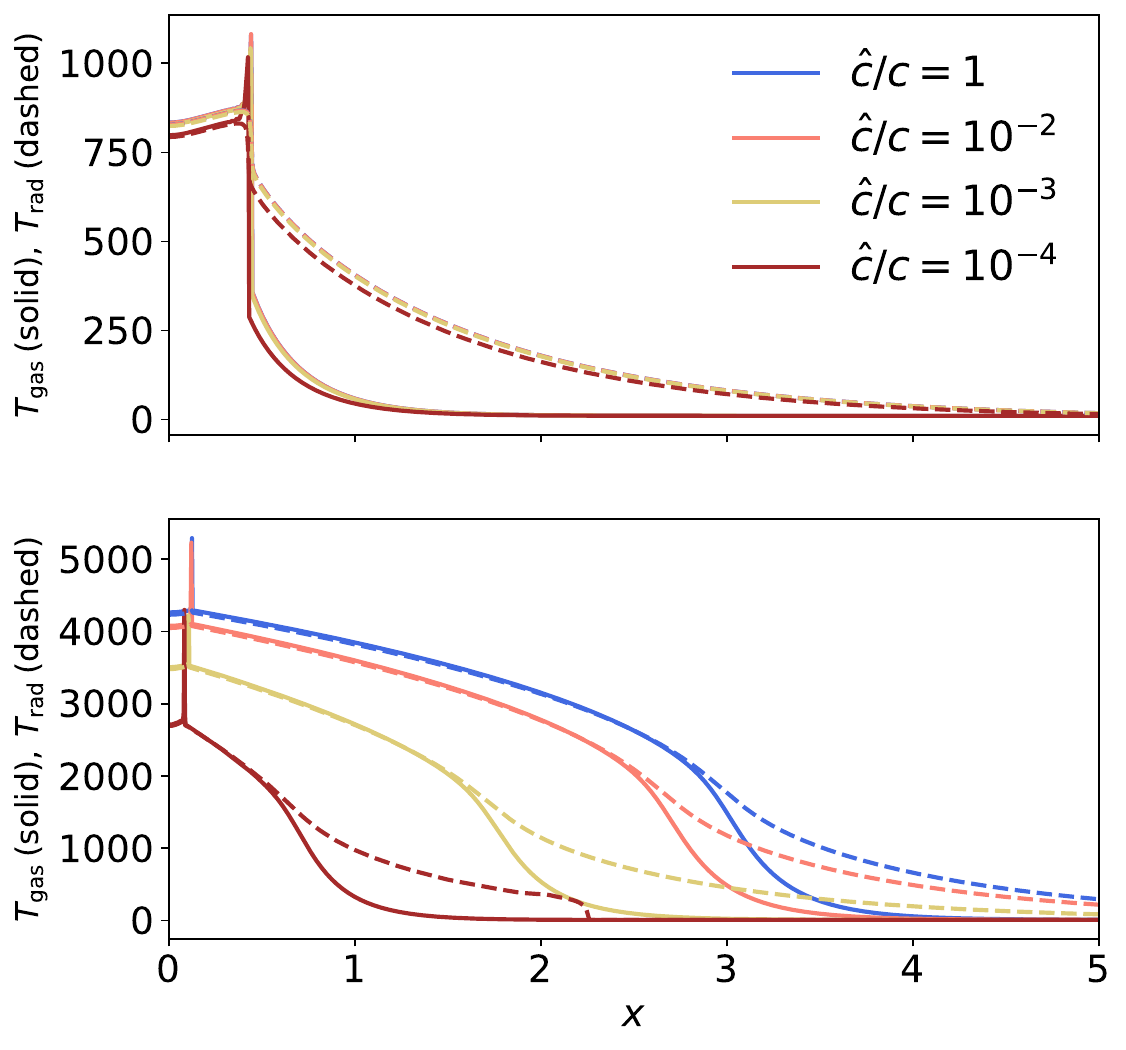}
    \caption{Gas temperature (solid lines) and radiation temperature (dashed lines) as a function of $x$ for the subcritical radiation shock (\textit{top}) and supercritical shock (\textit{bottom}) for different choices of reduced speed of light $\hat{c}/c=1,10^{-2},10^{-3}$, and $10^{-4}$ as blue, salmon, yellow, and brown lines, respectively.}
    \label{fig:shock}
\end{figure}

Radiative shocks differ from hydrodynamical shocks as radiation is able to redistribute the thermal energy deposited in the post-shock region, which is heated to a temperature of $T_2$, back into the pre-shock region. In an optically thick medium, the pre-shock region will rise to a temperature $T_-\le T_2$ before reaching the front. As a result, the temperature on the shocked side of the front is $T_+>T_2$ (see Fig. \ref{fig:shock}) and relax to $T_2$ away from the front. There are two types of radiative shocks depending on the strength of the shock: 1) subcritical shock, where the post-shock temperature $T_2$ is not high enough for radiation to dramatically affect the pre-shock region so that $T_-<T_2$, 2) supercritical shock, where $T_2$ is so large that the radiation escaping from the post-shock region is able to heat the precursor to $T_-=T_2$.

We perform two radiative shocks simulations, a subcritical one and a supercritical one, to test the dependence of our results with the reduced speed of light. Indeed, since radiative shocks are dynamic problems involving light-matter interaction, they provide a good illustration of how our choice of reduced speed of light can affect the final solution. As noted in Sect. \ref{sec:reduced_c}, the use of the reduced speed of light approximation introduces an artificial loss of energy such that
\begin{equation}
\Delta(E+E_r) = (1-\hat{c}/c)\Delta E_r,
\end{equation} 
which is particularly important in problems where the radiative energy changes suddenly such as in the case of radiative shocks. 

To initialize the simulation, we reproduced the setup of \cite{ensman1994}, \cite{hayes2003}, \cite{gonzalez2007}, and \cite{melon2021}. We initialized a 1D Cartesian grid extending from $x\in[0,7\times10^{10}\:\mathrm{cm}]$. We initialized a constant density of $\rho=7.78\times 10^{-10}\:\mathrm{g\:cm^{-3}}$ and a constant pressure and radiation field using a temperature of $T=10\:K$, $\mu=1$ and $\Gamma=7/5$. We used a constant absorption opacity of $\kappa=0.39$ and a null scattering opacity so that the domain is optically thick to absorption opacity with $\rho\kappa L\approx 20$. A rightward moving shock was created by setting $v_x<0$ and a reflective boundary condition on the left boundary. Our choice of $v_x$ determines the shock regime. We used $v_x=-6$ and $-20\:\mathrm{km\:s^{-1}}$ for the subcritical and supercritical regime, respectively. Finally, we used the HLL Riemann solver with the PLM $f$-preserving reconstruction.

We see in Fig. \ref{fig:shock} that in both cases for $\hat{c}/c=1$, we retrieve a very similar solution as \cite{gonzalez2007} and \cite{melon2021}, with $T_2=834\:K$, $T_-=358\:K,$ and $T_+=1080\:K$ for the subcritical shock and $T_2=4262\:K$ and $T_+=5314\:K$ for the supercritical shock. We also see that our choice of $\hat{c}/c$ critically affect the solution, especially for the supercritical shock. For the subcritical shock, we find $T_2=800\:K$ for $\hat{c}/c=10^{-4}$ quite close to the solution with $\hat{c}/c=1$. However, for the supercritical shock, we find $T_2=2720\:K$ for $\hat{c}/c=10^{-4}$ much lower than for $\hat{c}/c=1$. Interestingly, we see that a naive prediction of the maximum $\hat{c}$ we can use for the subcritical shock would lead to $\hat{c}\gg (|v_x|+c_s)\approx9\times10^{-5}$ while our departure from the $\hat{c}$ solution is already very large at $\hat{c}=10^{-3}$. However, as noted by \cite{melon2021}, this simple estimate does not take into account the timescale on which energy is injected into the system. We propose that a better estimate of $\hat{c}$ is given by $\max (E_r/|\rho v_x^3|)$, which gives the typical velocity at which radiation can escape in order to compensate for kinetic energy injection. We find that at the shock location of our simulation with $\hat{c}/c=1$, this estimate gives $\hat{c}\gg1230$ and $\hat{c}\gg 1$ for the subcritical and supercritical shocks. This is consistent with the fact that the subcritical shock solution departs from the reference solution only for $\hat{c}/c=10^{-4}$ while the supercritical shock departs from the reference solution for $\hat{c}/c=10$ already. This highlights the importance of examining all relevant timescales of the simulations and of performing convergence tests with different reduced speeds of light.

\subsection{Vertical diffusion}

We perform a test of radiative energy diffusion along the vertical extent of a disk under constant injection of viscous energy (see \citealt{melon2021}). With this test, we check the ability of our code to perform in the diffusion limit. We also tested the diffusion of an optically thick pulse (as in \citealt{melon2019}) but do not present it here as the two tests are testing similar features of the code.

We model a 1D vertical slice of a disk in hydrostatic equilibrium as 
\begin{equation}
\rho(x)=\rho_0\exp(-x^2/2H^2)+\rho_\mathrm{min},
\end{equation}
where $\rho_0=10^{-10}\:\mathrm{g\:cm^{-3}}$ is the midplane density, $\rho_\mathrm{min}=10^{-10}\rho_0$ is the density floor, $H=0.05 R$ is the pressure scale height and $x\in[-1,1]$ is the height of the disk. The pressure is initialized using a constant temperature $T_0=1000\:K$. We used a mean molecular weight $\mu=2.35$ and an adiabatic index $\Gamma=1.41$. We used a resolution of 201 points in $x$, an HLL solver with PLM $f$-preserving reconstruction, and a RK2 time-integrator. Finally, we set local thermodynamic equilibrium with $T=T_0$ at the boundaries for $E_r$ as well as zero-gradient boundary conditions on $F_r$. 

We set all velocity components to zero and set the mass and momentum fluxes to zero so as to keep the density and velocity structure fixed.  Hence, the energy is evolved only under the influence of radiative cooling and viscous heating defined as 
\begin{equation}
    S_E = \frac{9}{4}\alpha\Omega_Kc_s^2\rho,
\end{equation}
where $\alpha=10^{-3}$, $\Omega_K$ is the Keplerian angular velocity at 5 au, and $c_s$ is the isothermal sound speed.

\begin{figure}
    \centering
    \includegraphics[width=90mm]{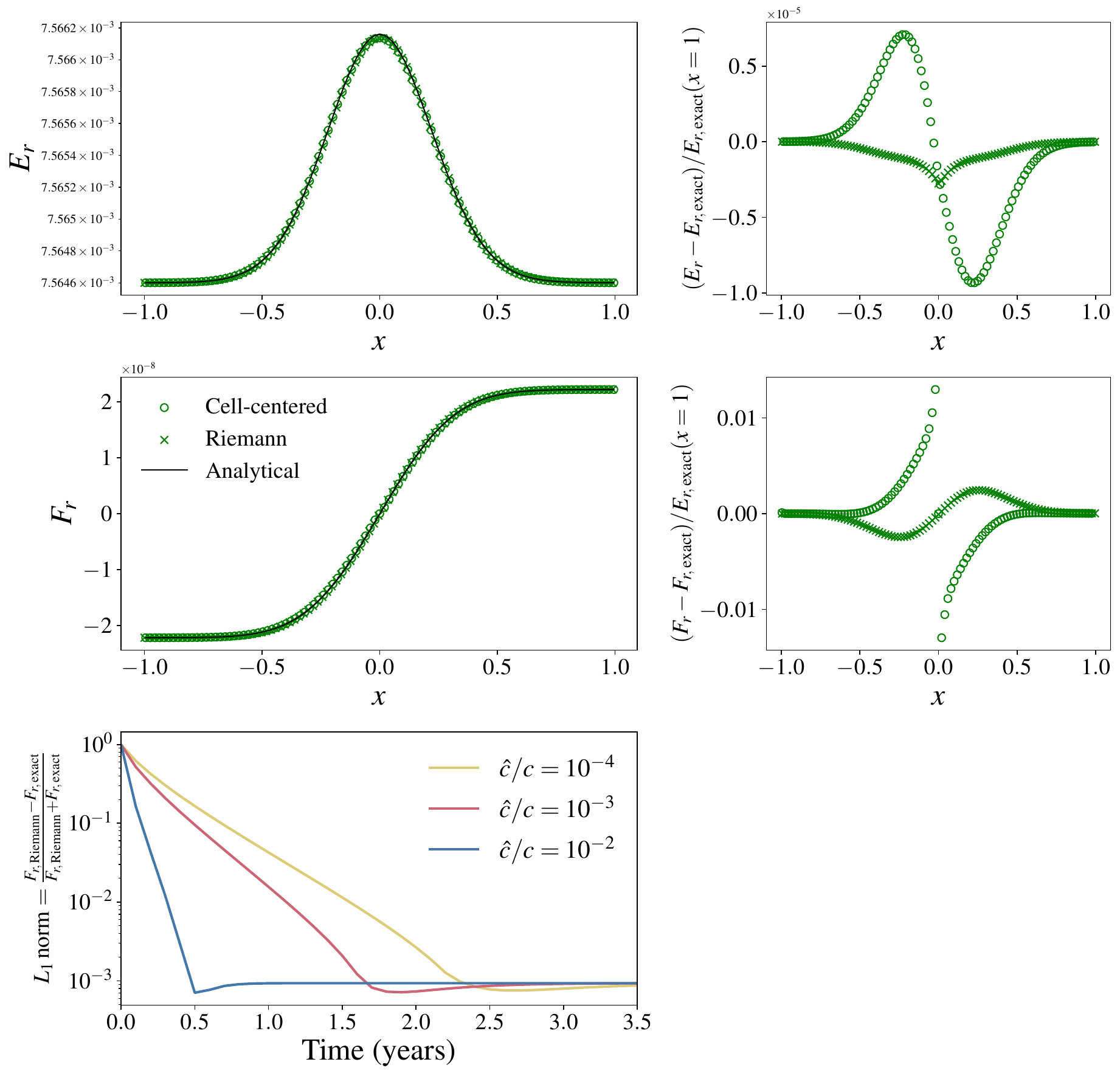}
    \caption{\textit{Top left} and \textit{center left}: Radiative energy and radiative flux as a function of $x$ for the vertical diffusion test. Empty circles show the cell-centered values. Crosses show the values of radiative flux from the Riemann solver and radiative energy reconstructed from the Riemann fluxes of energy and flux. The black line shows the analytical solution. \textit{Top right} and \textit{center right}: Difference between our simulation and the analytical solution. \textit{Bottom}: L1 relative norm for three different reduced speed of light $\hat{c}/c=10^{-2},10^{-3}$, and $10^{-4}$ as blue, red, and yellow lines, respectively.}
    \label{fig:diffusion}
\end{figure}

We plot in Fig. \ref{fig:diffusion} the semi-analytical solution to this problem, which we computed following the procedure of \cite{melon2021}, as well as solutions for $\hat{c}/c=1,\:10^{-2}$ and $10^{-4}$. We see that our solution is very close to the analytical solution for all $\hat{c}/c$. However, we find that find cell-centered variables always exhibit a flattening of the radiative flux near the origin that can be clearly seen in the right panels of Fig. \ref{fig:diffusion}. This flattening of the radiative flux is actually an artifact due to the reconstruction of cell-centered variables on the cell faces when solving the Riemann problem. Indeed, we plot the radiative flux from the Riemann problem and find that it does not show a flattening. We also plot in the bottom panel of Fig. \ref{fig:diffusion} the relative L1 norm error on the radiative flux as a function of time. We see that the final error on our solution is identical for all of our values of reduced speed of light and is around $10^{-3}$. The only difference between simulations using different reduced speed of light is the time they take to get to the final solution. We note that in this test, although we do inject energy all the time as in the radiative shock test, the characteristic time of energy injection is $\min(E_r/S_E)\approx 5\times10^7$ s, which is ten times slower than the typical light crossing timescale for $\hat{c}/c=10^{-4}$.

\subsection{Stellar irradiation}\label{sec:irradiation_test}
In this last test, we checked the ability of our code to handle irradiation of a passive disk by a central star  \citep{pascucci2004}. We defined our disk density as
\begin{equation}
\rho(R,z)=\rho_0\left(\frac{500\:\mathrm{AU}}{R}\right)\exp \left( -\frac{\pi}{4}\left(\frac{z}{h(r)}\right)^2\right),
\end{equation}
where $(R,z)=r(\cos \theta,\sin\theta)$ with $r\in[1,1000]$ AU and $\theta\in[0,\pi]$ and where $h(R)=125\:\mathrm{AU}\times(R/500\:\mathrm{AU})^{1.125}$. We used a spherical grid with 240 cells spaced logarithmically in $r$ and 100 cells spaced uniformly in $\theta$. We used outflowing boundary conditions in all directions and used an HLL solver with PLM reconstruction and a RK2 time integrator. We used a reduced speed of light of $\hat{c}/c=10^{-4}$ and ran the simulation for 10 years of physical time until the temperature reached equilibrium out to $R\approx300$ AU. We set all velocity components to zero and we cancel the fluxes of density, momentum, and energy so that the disk temperature is only changed by the irradiation source term and the absorption source term.

\begin{figure}
    \centering
    \includegraphics[width=90mm]{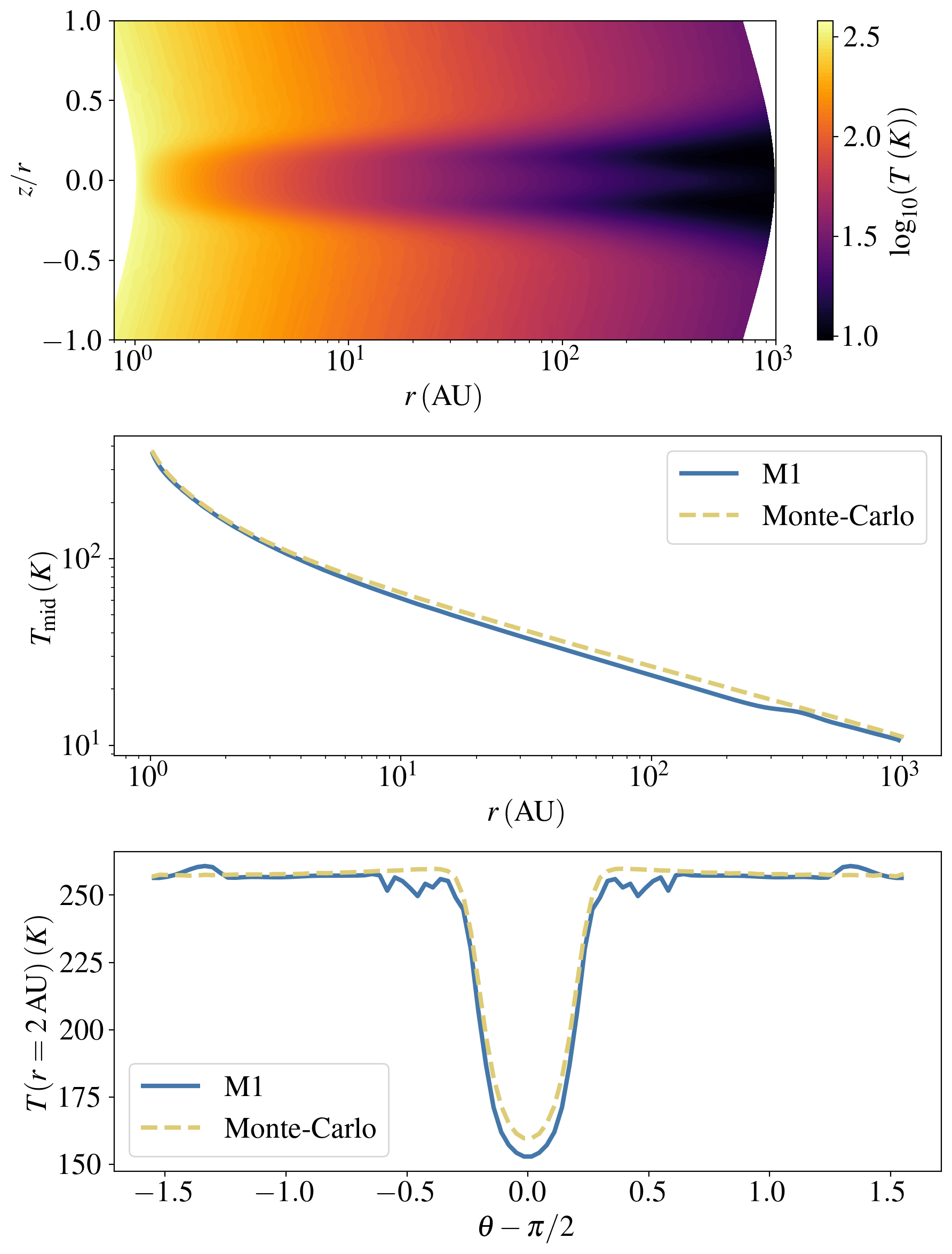}
    \caption{\textit{Top}: Color map of the gas temperature as a function of $r$ and $z/r$ for the stellar irradiation test. \textit{Middle} and \textit{bottom}: Comparison of the temperature for the Monte Carlo and M1 solution in the midplane as a function of $r$ (\textit{middle}) and at $r=2$ AU as a function of $\theta$ (\textit{bottom}).}
    \label{fig:stellar_irradiation}
\end{figure}

We computed the opacities for absorption and irradiation from the opacities provided by RADMC3D. We used the "astronomical silicate" opacity of \cite{draine2003}, with a grain size of $0.12\:\mu\mathrm{m}$ and a density of $3.6\:\mathrm{g\:cm^{-3}}$. We set $\rho_0=6.66\times10^{-17}\:\mathrm{g\:cm^{-3}}$ so that the optical depth of a ray at $550$ nm going through the midplane experiences an optical depth of $\tau=100$. From these opacities as a function of wavelength, we pre-computed 1D Planck and Rosseland opacity tables as a function of temperature that we use to compute the opacity by interpolation in each cell. We neglect scattering by setting $\sigma=0$.

The irradiation flux is given by 
\begin{equation}\label{eq:irr_flux_stellar}
\textbf{F}_\mathrm{irr}(r,\theta)=\pi\left( \frac{R_s}{r} \right)^2 \int_{\nu_\mathrm{min}}^{\nu_\mathrm{max}}d\nu B_\nu(T_s)e^{-\tau_\mathrm{irr}(r,\theta,\nu)}\textbf{r},
\end{equation}
where $R_s=R_\odot$ is the radius of the central star,$\nu_\mathrm{min}=1.5\times 10^{11}$ Hz and $\nu_\mathrm{max}=1.5\times 10^{15}$ Hz, $B_\nu$ is the Planck function and $T_s=5800$ K is the temperature of the central star. We define the optical depth seen by the irradiation flux as 
\begin{equation}
    \tau_\mathrm{irr}(r,\theta,\nu) = \int_{r_\mathrm{in}}^r \kappa(\nu)\rho(r',\theta,\nu)dr'.
\end{equation}
In practice, we pre-compute tables of the integral on frequency in Eq. \ref{eq:irr_flux_stellar} as a function of $\tau_\mathrm{irr}$ from which we interpolate the flux in each cell. 

\begin{figure}
    \centering
    \includegraphics[width=90mm]{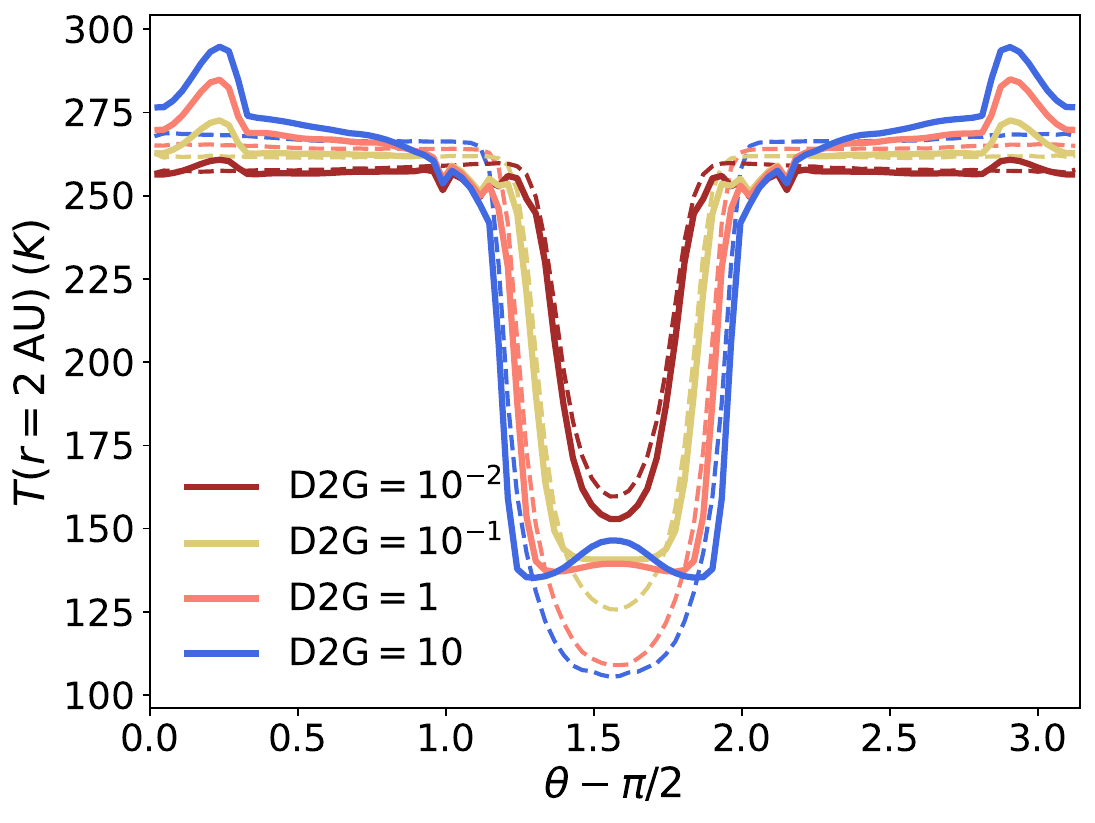}
    \caption{Temperature at $r=2$ AU as a function of $\theta$ for four different dust-to-gas ratios of $10^{-2},10^{-1},1,$ and 10, shown as brown, yellow, salmon, and blue lines, respectively. Dashed lines show the results from the Monte Carlo simulations, and solid lines show the results from the M1 simulations.}
    \label{fig:stellar_irradiation_failure}
\end{figure}

We show in Fig. \ref{fig:stellar_irradiation} a color map of the temperature as well as a cut in the midplane as a function of radius and a cut at a radius of 2 AU as a function of $\theta$. We retrieve the typical structure of an irradiated disk, with hot upper layers and a cold disk midplane where stellar irradiation gets absorbed as it propagates radially outward. We compare our solution with a multifrequency multidimensional Monte Carlo simulation done with RADMC3D for the same grid and the same opacity tables using $10^8$ photons. We see that the agreement is quite good between our simulation and the Monte Carlo simulation. We typically overestimate the temperature near the pole, because of beam crossing at the axis, and we underestimate the temperature in the midplane, but these deviations between the two methods do not exceed $5\%$. Nonetheless, we find that at higher optical depth the disagreement between the two method gets worse. We plot in Fig. \ref{fig:stellar_irradiation_failure} cuts of the temperature at 2 AU for increasing dust to gas ratios (so increasing optical depth). We see that as we increase the optical depth, the midplane temperature flattens and eventually gets a "Mexican hat" shape. This problem was raised in \cite{melon2025} where it was attributed to beam crossing at the midplane. In \cite{melon2025}, the authors proposed a modified M1 method, called the half moment method, where the equations are integrated over hemispheres instead of the full solid angle when computing the moment of the radiative transfer equations. However, implementing this method is outside the scope of this paper.

\section{Performance}\label{sec:perf}
To test the performance of our radiative version of \idefix{}, we perform the pulse test described in Sect. \ref{sec:pulse} on a 3D Cartesian grid. We let the simulation run for a thousand cycles without performing any outputs. We use the HLL solver for radiation and the hydrodynamics with the PLM $f$-preserving scheme and RK2 time-integration. When we increase the number of nodes, we increase proportionally the size of the domain as well as the number of points in the simulation. In this way, we maintain a constant time step in the simulation and so a fixed number of cycles. 

\begin{figure}
    \centering
    \includegraphics[width=90mm]{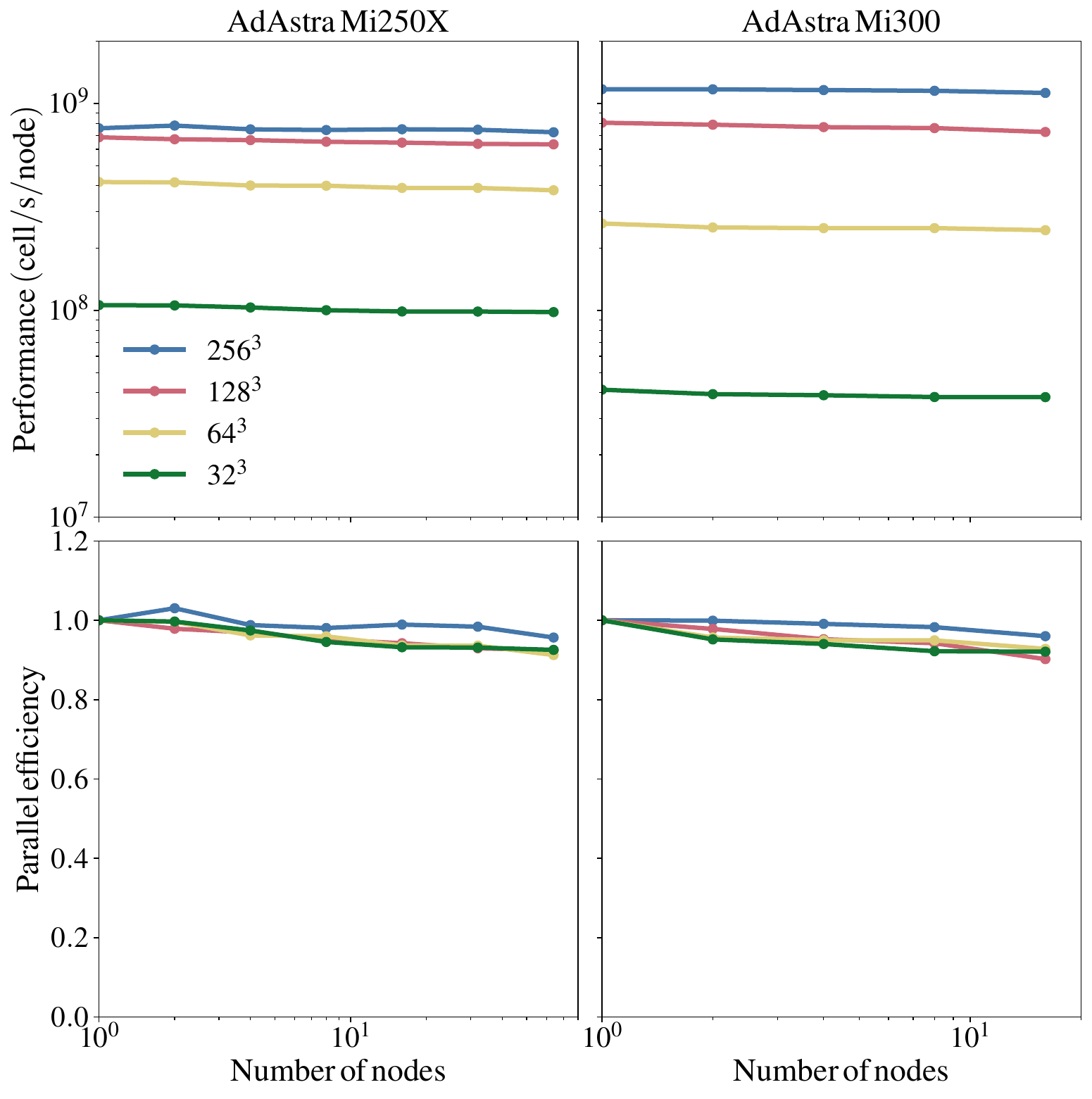}
    \caption{\textit{Top}: Performance-in-cell update per second per cell on AdAstra's Mi250x and Mi300 as a function of nodes. \textit{Bottom}: Weak scaling efficiency.}
    \label{fig:perf}
\end{figure}

We measure the performance of \idefix{} on the French supercomputer AdAstra on two types of GPUs, AMD Mi250X and AMD Mi300. We plot in the top panels of Fig. \ref{fig:perf} the number of cell updates per second per node as a function of the number of nodes. We see that for the largest sub-domains of $256^3$, the performance goes as high as $7.6\times10^8$ cell updates per second per node on Mi250X and $1.2\times10^9$ cell updates per second per node on Mi300. This compares to $1.24\times10^9$ cell updates per second per node on Mi250X for a test run on MHD (see \citealt{lesur2023}). This shows that the radiative transfer scheme is roughly 1.6 times more expensive than the MHD scheme, because of the complexity of the operations needed to compute the closure in the former case. Interestingly, we see that although the performance on the Mi300 GPUs is better than on the Mi250X GPUs for sub-domains of $256^3$, it is the contrary for sub-domains of $64^3$ and $32^3$. This behavior arises because \idefix{} employs two MPI processes per GPU on the Mi250 (one process per graphic compute die), effectively increasing the domain size per GPU. As a result, the Mi250 can better hide memory access latency compared to the Mi300 for small domain sizes.

We also plot the weak scaling of our simulations in the lower panels of Fig. \ref{fig:perf}. We see that on both Mi250x and Mi300 our radiative version of \idefix{} performs with almost perfect scaling, staying higher than $90\%$ even when running on $64$ nodes. This is consistent with the test presented on the MHD version of \idefix{} and is quite natural given how close is the implementation of the radiative module and the hydrodynamic module. 

\section{Conclusion}\label{sec:conclusion}
In this paper, we have presented the new radiative transfer module of \idefix{}. We used the M1 approximation, effectively treating radiation as a fluid, which is evolved using a high-order finite-volume Godunov method. We solved for the hyperbolic part of the equations using an explicit method with radiation Riemann solvers. To compute the reconstructed inter-cell fluxes, we propose an improved reconstruction scheme that provides more accurate results in the free-streaming limit. Because the scheme is nonrelativistic, we used a reduced speed of light to reduce the difference between the hydrodynamic and radiative timescales. The source terms are treated implicitly by inverting a matrix of $(n+1)^2$, where $n$ is the number of frequency groups. For now, the scheme uses only one frequency group, but we intend to use several in future work. Radiation can be used on Cartesian, cylindrical and spherical grids that are uniform or stretched and in multiple dimensions. The scheme is also compatible with the MHD module. 

The radiative transfer version of \idefix{} performs well, with as many as $7.6\times10^8$ cell updates per second per node on an AMD Mi250x on AdAstra. It keeps a weak scaling efficiency of $90\%$ up to 64 nodes. As such, radiation only costs $1.6$ times an MHD simulation. 

\begin{acknowledgements}
The authors acknowledge support from the European Research Council (ERC) under the European Union Horizon 2020 research and innovation program (Grant agreement No. 815559 (MHDiscs)). This work was supported by the MHD@Exascale project (reference 22-EXOR-0015) of PEPR Origins (PI: Morbidelli). This project was provided with computing HPC and storage resources by GENCI at CINES thanks to the grant 2025-A0180402231 on the supercomputer Adastra Mi250 and Mi300 partitions.
\end{acknowledgements}

 \bibliographystyle{aa} 
 \bibliography{biblio}

\begin{thebibliography}{42}
\expandafter\ifx\csname natexlab\endcsname\relax\def\natexlab#1{#1}\fi

\bibitem[{{Colella} \& {Sekora}(2008)}]{colella2008}
{Colella}, P. \& {Sekora}, M.~D. 2008, Journal of Computational Physics, 227, 7069

\bibitem[{Colella \& Woodward(1984)}]{colella1984}
Colella, P. \& Woodward, P.~R. 1984, Journal of Computational Physics, 54, 174

\bibitem[{{Commer{\c{c}}on} {et~al.}(2011){Commer{\c{c}}on}, {Teyssier}, {Audit}, {Hennebelle}, \& {Chabrier}}]{commercon2011}
{Commer{\c{c}}on}, B., {Teyssier}, R., {Audit}, E., {Hennebelle}, P., \& {Chabrier}, G. 2011, \aap, 529, A35

\bibitem[{{Dexter} {et~al.}(2021){Dexter}, {Scepi}, \& {Begelman}}]{dexter2021}
{Dexter}, J., {Scepi}, N., \& {Begelman}, M.~C. 2021, \apjl, 919, L20

\bibitem[{{Draine}(2003)}]{draine2003}
{Draine}, B.~T. 2003, \apj, 598, 1017

\bibitem[{{Ensman}(1994)}]{ensman1994}
{Ensman}, L. 1994, \apj, 424, 275

\bibitem[{{Flock} {et~al.}(2017){Flock}, {Nelson}, {Turner}, {Bertrang}, {Carrasco-Gonz{\'a}lez}, {Henning}, {Lyra}, \& {Teague}}]{flock2017}
{Flock}, M., {Nelson}, R.~P., {Turner}, N.~J., {et~al.} 2017, \apj, 850, 131

\bibitem[{{Foucart}(2018)}]{foucart2018}
{Foucart}, F. 2018, \mnras, 475, 4186

\bibitem[{{Foucart}(2023)}]{foucart2023}
{Foucart}, F. 2023, Living Reviews in Computational Astrophysics, 9, 1

\bibitem[{{Gnedin} \& {Abel}(2001)}]{Gnedin2001}
{Gnedin}, N.~Y. \& {Abel}, T. 2001, \na, 6, 437

\bibitem[{{Gonz{\'a}lez} {et~al.}(2007){Gonz{\'a}lez}, {Audit}, \& {Huynh}}]{gonzalez2007}
{Gonz{\'a}lez}, M., {Audit}, E., \& {Huynh}, P. 2007, \aap, 464, 429

\bibitem[{{Hayes} \& {Norman}(2003)}]{hayes2003}
{Hayes}, J.~C. \& {Norman}, M.~L. 2003, \apjs, 147, 197

\bibitem[{{Hirose} {et~al.}(2009){Hirose}, {Blaes}, \& {Krolik}}]{hirose2009}
{Hirose}, S., {Blaes}, O., \& {Krolik}, J.~H. 2009, \apj, 704, 781

\bibitem[{{Jiang} {et~al.}(2025){Jiang}, {Blaes}, {Kaul}, \& {Zhang}}]{jiang2025}
{Jiang}, Y.-F., {Blaes}, O., {Kaul}, I., \& {Zhang}, L. 2025, \apj, 988, 43

\bibitem[{{Jiang} {et~al.}(2012){Jiang}, {Stone}, \& {Davis}}]{davis2012}
{Jiang}, Y.-F., {Stone}, J.~M., \& {Davis}, S.~W. 2012, \apjs, 199, 14

\bibitem[{{Jiang} {et~al.}(2019){Jiang}, {Stone}, \& {Davis}}]{jiang2019}
{Jiang}, Y.-F., {Stone}, J.~M., \& {Davis}, S.~W. 2019, \apj, 880, 67

\bibitem[{Kudritzki \& Puls(2000)}]{kudritzki2000}
Kudritzki, R.-P. \& Puls, J. 2000, Annual Review of Astronomy and Astrophysics, 38, 613

\bibitem[{{Lesur} {et~al.}(2023){Lesur}, {Baghdadi}, {Wafflard-Fernandez}, {Mauxion}, {Robert}, \& {Van den Bossche}}]{lesur2023}
{Lesur}, G.~R.~J., {Baghdadi}, S., {Wafflard-Fernandez}, G., {et~al.} 2023, \aap, 677, A9

\bibitem[{LeVeque(2002)}]{leveque2002}
LeVeque, R.~J. 2002, Finite volume methods for hyperbolic problems, Vol.~31 (Cambridge university press)

\bibitem[{{Levermore}(1984)}]{levermore1984}
{Levermore}, C.~D. 1984, \jqsrt, 31, 149

\bibitem[{{Liska} {et~al.}(2022){Liska}, {Musoke}, {Tchekhovskoy}, {Porth}, \& {Beloborodov}}]{liska2022}
{Liska}, M.~T.~P., {Musoke}, G., {Tchekhovskoy}, A., {Porth}, O., \& {Beloborodov}, A.~M. 2022, \apjl, 935, L1

\bibitem[{{McKinney} {et~al.}(2014){McKinney}, {Tchekhovskoy}, {Sadowski}, \& {Narayan}}]{mckinney2014}
{McKinney}, J.~C., {Tchekhovskoy}, A., {Sadowski}, A., \& {Narayan}, R. 2014, \mnras, 441, 3177

\bibitem[{{Melon Fuksman} {et~al.}(2025){Melon Fuksman}, {Flock}, {Klahr}, {Mattia}, \& {Muley}}]{melon2025}
{Melon Fuksman}, D., {Flock}, M., {Klahr}, H., {Mattia}, G., \& {Muley}, D. 2025, \aap, 701, A97

\bibitem[{{Melon Fuksman} {et~al.}(2021){Melon Fuksman}, {Klahr}, {Flock}, \& {Mignone}}]{melon2021}
{Melon Fuksman}, J.~D., {Klahr}, H., {Flock}, M., \& {Mignone}, A. 2021, \apj, 906, 78

\bibitem[{{Melon Fuksman} \& {Mignone}(2019)}]{melon2019}
{Melon Fuksman}, J.~D. \& {Mignone}, A. 2019, \apjs, 242, 20

\bibitem[{Mezzacappa {et~al.}(2020)Mezzacappa, Endeve, Messer, \& Bruenn}]{mezzacappa2020}
Mezzacappa, A., Endeve, E., Messer, O.~B., \& Bruenn, S.~W. 2020, Living Reviews in Computational Astrophysics, 6, 4

\bibitem[{Mihalas \& Mihalas(2013)}]{mihalas2013}
Mihalas, D. \& Mihalas, B.~W. 2013, Foundations of radiation hydrodynamics (Courier Corporation)

\bibitem[{{Miller} {et~al.}(2019){Miller}, {Ryan}, {Dolence}, {Burrows}, {Fontes}, {Fryer}, {Korobkin}, {Lippuner}, {Mumpower}, \& {Wollaeger}}]{miller2019}
{Miller}, J.~M., {Ryan}, B.~R., {Dolence}, J.~C., {et~al.} 2019, \prd, 100, 023008

\bibitem[{{Miller} {et~al.}(2020){Miller}, {Sprouse}, {Fryer}, {Ryan}, {Dolence}, {Mumpower}, \& {Surman}}]{miller2020}
{Miller}, J.~M., {Sprouse}, T.~M., {Fryer}, C.~L., {et~al.} 2020, \apj, 902, 66

\bibitem[{{Minerbo}(1978)}]{minerbo1978}
{Minerbo}, G.~N. 1978, \jqsrt, 20, 541

\bibitem[{{Murray} {et~al.}(2010){Murray}, {Quataert}, \& {Thompson}}]{norman2010}
{Murray}, N., {Quataert}, E., \& {Thompson}, T.~A. 2010, \apj, 709, 191

\bibitem[{{Pascucci} {et~al.}(2004){Pascucci}, {Wolf}, {Steinacker}, {Dullemond}, {Henning}, {Niccolini}, {Woitke}, \& {Lopez}}]{pascucci2004}
{Pascucci}, I., {Wolf}, S., {Steinacker}, J., {et~al.} 2004, \aap, 417, 793

\bibitem[{{Proga} {et~al.}(1998){Proga}, {Stone}, \& {Drew}}]{proga1998}
{Proga}, D., {Stone}, J.~M., \& {Drew}, J.~E. 1998, \mnras, 295, 595

\bibitem[{{Richling} {et~al.}(2001){Richling}, {Meink{\"o}hn}, {Kryzhevoi}, \& {Kanschat}}]{richling2001}
{Richling}, S., {Meink{\"o}hn}, E., {Kryzhevoi}, N., \& {Kanschat}, G. 2001, \aap, 380, 776

\bibitem[{{Ripoll} {et~al.}(2001){Ripoll}, {Dubroca}, \& E.}]{ripoll2001}
{Ripoll}, J.-F., {Dubroca}, B., \& E., D. 2001, Combustion Theory and Modelling, 5, 261

\bibitem[{{Rosdahl} {et~al.}(2013){Rosdahl}, {Blaizot}, {Aubert}, {Stranex}, \& {Teyssier}}]{rosdahl2013}
{Rosdahl}, J., {Blaizot}, J., {Aubert}, D., {Stranex}, T., \& {Teyssier}, R. 2013, \mnras, 436, 2188

\bibitem[{{Rosdahl} \& {Teyssier}(2015)}]{rosdahl2015}
{Rosdahl}, J. \& {Teyssier}, R. 2015, \mnras, 449, 4380

\bibitem[{{Roth} {et~al.}(2025){Roth}, {Anninos}, {Fragile}, \& {Pickrel}}]{roth2025}
{Roth}, N., {Anninos}, P., {Fragile}, P.~C., \& {Pickrel}, D. 2025, \apj, 981, 144

\bibitem[{{S{\k{a}}dowski} {et~al.}(2013){S{\k{a}}dowski}, {Narayan}, {Tchekhovskoy}, \& {Zhu}}]{sadowski2013}
{S{\k{a}}dowski}, A., {Narayan}, R., {Tchekhovskoy}, A., \& {Zhu}, Y. 2013, \mnras, 429, 3533

\bibitem[{{Skinner} \& {Ostriker}(2013)}]{skinner2013}
{Skinner}, M.~A. \& {Ostriker}, E.~C. 2013, \apjs, 206, 21

\bibitem[{{Thompson} {et~al.}(2005){Thompson}, {Quataert}, \& {Murray}}]{thompson2005}
{Thompson}, T.~A., {Quataert}, E., \& {Murray}, N. 2005, \apj, 630, 167

\bibitem[{{Čada} \& {Torrilhon}(2009)}]{cada2009}
{Čada}, M. \& {Torrilhon}, M. 2009, Journal of Computational Physics, 228, 4118

\end{thebibliography}

\end{document}